\documentclass[aps,prl,showpacs,twocolumn,superscriptaddress, floatfix, tightenlines, amsmath, amssymb]{revtex4-1}
\usepackage[pdftex]{graphicx} %[dvips]

\graphicspath{{figures/}}
\usepackage{natbib}
\usepackage{booktabs}
\usepackage{xcolor}
\usepackage{url}
\usepackage{amsmath}
\usepackage{bm}
\usepackage[colorlinks=true, citecolor=blue, linkcolor=blue, urlcolor=blue]{hyperref}
\usepackage[inline]{enumitem}

\newcommand{\AgIr}{Ag$_3$LiIr$_2$O$_6$} 
\newcommand{\Li}{$\alpha$-Li$_2$IrO$_3$}
\newcommand{\eloss}{$E_{\mathrm{loss}}$}
\newcommand{\LS}{\langle L\cdot S\rangle}
\newcommand{\Lz}{\langle L_z \rangle}
\newcommand{\Sz}{\langle S_z \rangle}
\newcommand{\Tz}{\langle T_z \rangle}
\newcommand{\ub}{\mu_B}

\begin{document}

\title{Nearly itinerant electronic groundstate in the intercalated honeycomb iridate \AgIr{}}
\author{A. de la Torre}
\affiliation{Department of Physics, Brown University, Providence, Rhode Island 02912, United States}
\author{B. Zager}
\affiliation{Department of Physics, Brown University, Providence, Rhode Island 02912, United States}
\author{F. Bahrami}
\affiliation{Department of Physics, Boston College, Chestnut Hill, MA 02467, USA}
\author{M. DiScala}
\affiliation{Department of Physics, Brown University, Providence, Rhode Island 02912, United States}
\author{J. R. Chamorro}
\affiliation{Department of Chemistry, The Johns Hopkins University, Baltimore, MD, USA}
\affiliation{Institute for Quantum Matter, The Department of Physics and Astronomy, Johns Hopkins University, Baltimore, MD, USA}
\author{M. H. Upton}
\affiliation{Advanced Photon Source, Argonne National Laboratory, Argonne, Illinois 60439, USA}
\author{G. Fabbris}
\affiliation{Advanced Photon Source, Argonne National Laboratory, Argonne, Illinois 60439, USA}
\author{D. Haskel}
\affiliation{Advanced Photon Source, Argonne National Laboratory, Argonne, Illinois 60439, USA}
\author{D. Casa}
\affiliation{Advanced Photon Source, Argonne National Laboratory, Argonne, Illinois 60439, USA}
\author{T. M. McQueen}
\affiliation{Department of Chemistry, The Johns Hopkins University, Baltimore, MD, USA}
\affiliation{Institute for Quantum Matter, The Department of Physics and Astronomy, Johns Hopkins University, Baltimore, MD, USA}
\affiliation{Department of Materials Science and Engineering, The Johns Hopkins University, Baltimore, MD, USA}
\author{F. Tafti}
\affiliation{Department of Physics, Boston College, Chestnut Hill, MA 02467, USA}
\author{K. W. Plumb}
\email{kemp\_plumb@brown.edu}
\affiliation{Department of Physics, Brown University, Providence, Rhode Island 02912, United States}
\date{\today}

\begin{abstract}
We use x-ray spectroscopy at Ir L$_3$/L$_2$ absorption edge to study powder samples of the intercalated honeycomb magnet \AgIr{}. Based on x-ray absorption and resonant inelastic x-ray scattering measurements, and exact diagonalization calculations including next-neighbour Ir-Ir electron hoping integrals, we argue that the intercalation of Ag atoms results in a nearly itinerant electronic structure with enhanced Ir-O hybridization. As a result of the departure from the local relativistic $j_{\rm eff}\! = \!1/2$ state,  we find that the relative orbital contribution to the magnetic moment is increased, and the magnetization density is spatially extended and asymmetric. Our results confirm the importance of metal - ligand hybridazation in the magnetism of transition metal oxides and provide empirical guidance for understanding the collective magnetism in intercalated honeycomb iridates.
\end{abstract}
\maketitle
%--------------------- intro-----------------------------------
Recognition that the Kitaev model, an exactly soluble quantum spin liquid, may have material realization in heavy transition metal oxides has driven significant research for the past decade  \cite{PhysRevLett.102.017205}. The emergence of this effective quantum compass model depends on a hierarchy of crystal field, spin-orbit coupling, and electronic correlations that act to generate a relativistic atomic orbital basis with  $j_{\rm eff}\!=\!1/2$ effective angular moments. For insulators with edge sharing octahedra coordinating $j_{\rm eff}\!=\!1/2$ ions, isotropic Heisenberg exchange interactions nearly vanish and the magnetism is dominated by spatially anisotropic Kitaev exchange \cite{PhysRevLett.102.017205}. On the other hand, {\it ab initio} calculations find that for many model $j_{\rm eff}\!=\!1/2$ materials, kinetic energy can promote Ir-Ir covalency and the formation of a delocalized quasimolecular orbital (QMO) state \cite{PhysRevLett.109.197201, PhysRevB.88.035107}. Kitaev models are not obviously applicable in such covalent materials and little is known about the collective magnetism in the itinerant limit. 

In the context of Heisenberg-Kitaev magnetism, there has been intensive research on honeycomb iridates \cite{PhysRevLett.105.027204, PhysRevB.82.064412, PhysRevLett.108.127203} and  $\alpha$-RuCl$_3$ \cite{PhysRevB.90.041112, Banerjee1055, Banerjee2016}. These compounds have strong Kitaev exchange in addition to Heisenberg and pseudo-dipolar interactions that result in magnetic order \cite{PhysRevLett.112.077204, rau2014trigonal}. A successful strategy to reduce N\'eel temperatures has been to synthesize new versions of these compounds by substitution on the  alkali site through intercalation between the honeycomb planes, for example, \AgIr{} and H$_3$LiIr$_3$O$_6$ \cite{Abramchuk2017, Kitagawa2018, PhysRevLett.123.237203}. However, the $j_{\rm eff}\!=\!1/2$ state is fragile against structural details \cite{PhysRevB.88.035115, Clancy2018} while bond disorder and off-stoichiometries can inhibit magnetic order \cite{PhysRevLett.121.197203,PhysRevLett.121.247202}. An empirical understanding of the influence of chemical substitutions on this state is lacking and there does not yet exist a sound starting point to understand collective magnetism in these intercalated materials.

\begin{figure}[!t]
  \begin{center}
    \includegraphics{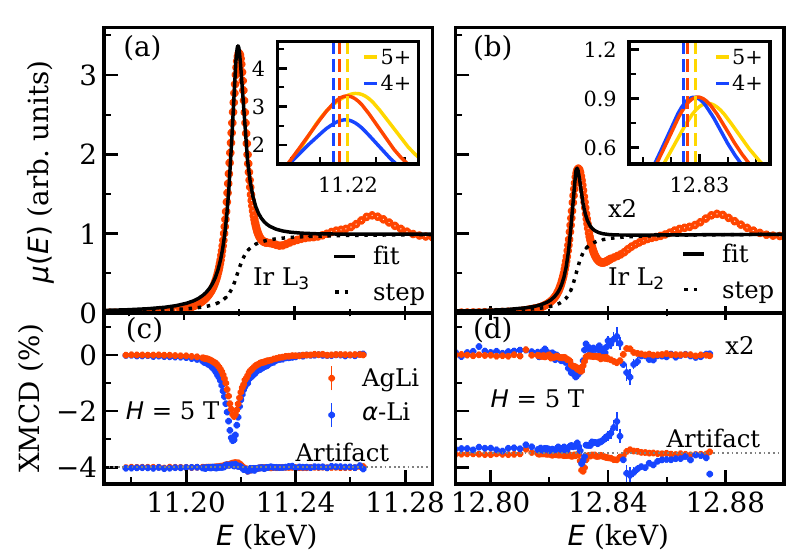}
  \caption{(a)-(b) Ir L$_3$ and  L$_2$ edge XAS at $T$=300~K. Black line is a fit, as explained in the main text. Insets in (a) and (b) detail L$_3$ and L$_2$  white lines for \AgIr{}, Li$_2$IrO$_3$ (Ir$^{4+}$), and Sr$_3$CaIr$_2$O$_9 $(Ir$^{5+}$). Dashed lines mark the inflection points. (c)-(d) $H = 5$~T and $T = 1.6$~K XMCD data for {\AgIr} and {\Li}. The magnitude of the spurious contributions due to experimental-artifact is shown. XAS and XMCD data at the L$_2$ edge is scaled by a factor of two for clarity.}
  \label{fig:Absorption}
  \end{center}
\end{figure}

In this letter, we use x-ray spectroscopies to show that the topochemical exchange of interlayer Li atoms with Ag in \AgIr{} \cite{bahrami2020effect,SM}, enhances Ir-O hybridization and fundamentally alters the magnetism. X-ray absorption reveals that the magnetism in \AgIr{} is characterized by an asymmetric spin density with strong spin-orbit coupling and a larger orbital component than the parent compound \Li{}. Resonant Inelastic X-ray Scattering (RIXS) spectra probing the Ir electronic structure of \AgIr{} is captured by incorporating Ir-O hybridization, demonstrating that the local $j_{\rm eff}\!=\!1/2$ picture is not a valid basis. \AgIr{} must be understood as a new type of nearly itinerant model quantum magnet. We posit that similar effects may be at play in other intercalated honeycomb iridates. Our results provide an empirical foundation to develop suitable effective Hamiltonians in these next-generation frustrated magnets. 

%--------------------- Methods ----------------------------------------------
High crystalline quality powder samples of \AgIr{}, $T_N = 8$~K, and \Li{} were from the same clean batch as in \cite{bahrami2020effect}. Samples of Sr$_3$CaIr$_2$O$_9$ were prepared as described in \cite{C5DT03188E}. RIXS was performed at Sector 27 (MERIX) at the Advanced Photon Source (APS) of the Argonne National Laboratory  \cite{SHVYDKO2013140}. We used a horizontal $2\theta\! =\! 90^{\circ}$ scattering geometry, with a 2 m radius spherically diced Si(844) analyzer to give a 33~meV overall energy resolution (full width at half maximum, FWHM). Powder diffraction was conducted at APS 11-ID-B using wavelength of 0.2115~\AA{} and sample to detector distance of 167.335 mm, calibrated with a ceria standard. The 2D diffraction data were integrated using GSAS-II and then corrected and normalized to obtain the atomic pair distribution function (PDF), $G(r)$ using PDFgetX3 \cite{PDFGetX3}. X-ray absorption near edge structure (XANES), far edge (EXAFS), and x-ray magnetic circular dichroism (XMCD) were performed in transmission at APS 4-ID-D. XAS and EXAFS data were analyzed using the Larch \cite{Newville_2013} and FEFF \cite{REHR2009548} software packages. Magnetic field and temperature were controlled using a 6.5~T LHe cooled magnet system. XMCD circular polarization was generated using a 500 $\mu$m diamond phase plate. 
%--------------------------------------------------------------------------------------------------
\begin{figure}[!t]
  \begin{center}
    \includegraphics{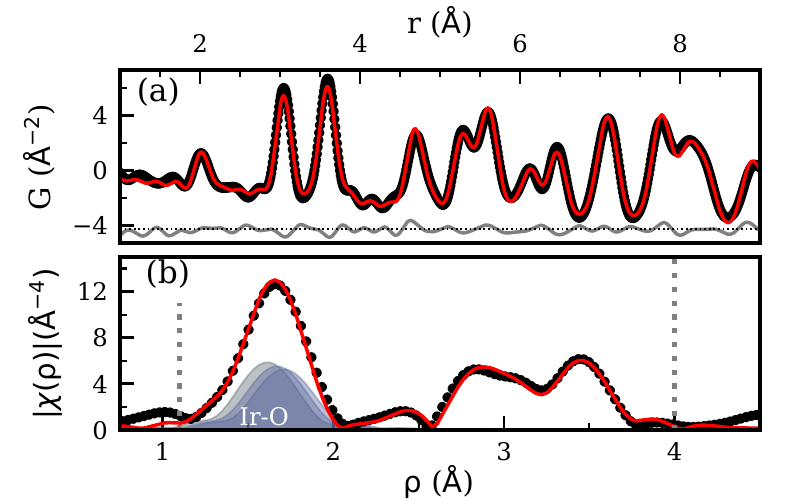}
  \caption{(a) Atomic PDF at $T\! = \! 300$~K, red line is a refinement in the range r = 1.2--12 \AA{} and grey line is difference between refinement and data. (b) Magnitude of the Fourier transform of  $T\! = \! 300$~K EXAFS oscillations and fit (red line). Shaded grey regions indicate the contribution from Ir-O bonds.}
  \label{fig:structure}
  \end{center}
\end{figure}

In Fig.~\ref{fig:Absorption} (a) and (b) we show the normalized absorption intensity at the Ir L$_3$ and L$_2$ edge respectively. Both the relative position of the $L_3$ white line inflection points and the L-edge sum rules provide quantitative information on the oxidation state of a transition metal ion \cite{PhysRevB.42.5459,PhysRevB.86.195131,PhysRevB.91.214433}. To extract this information, we fit the normalized data to an arctangent step and Lorentzian peak \cite{PhysRevB.97.035106}. The results of our analysis are summarized in Table~\ref{tab:XAS}. The inset of Fig.~\ref{fig:Absorption} shows a direct comparison of the white line measured in {\AgIr}, {\Li} ($5d^5$), and Sr$_3$CaIr$_2$O$_9$ ($5d^4$) \cite{C5DT03188E}. In \AgIr{}, we find $\langle n_h \rangle \!=\! 5.5(2)$ by averaging the results obtained from white line inflection points and integrated intensities. This deviates from $\langle n_h \rangle \!=\! 5$ expected for a localized $j_{\rm eff}\!=\!1/2$ state \cite{SM}. We also find a branching ratio, $BR\! =\! I_{L_3}/I_{L_2}\!=\!5.7(4)$ that is comparable to that of {\Li}; confirming that spin-orbit coupling is a dominant energy scale in \AgIr{} \cite{PhysRevLett.60.1977, PhysRevB.88.035107,PhysRevB.89.081104}. 

If the hole concentration extracted from the XAS analysis is interpreted to arise from a Ir$^{4+}$/Ir$^{5+}$ mixture, it implies that $50\%$ of the Ir are in a non-magnetic $j \!=\! 0$ state. Such a large fraction of Ir$^{5+}$ is not consistent with chemical analysis or the measured paramagnetic moment of $\mu_{\mathrm{eff}} \!=\! 1.87 \mu_{\mathrm{B}}$/Ir \cite{bahrami2020effect} and our XAS measurements at the Ag $L_3$-edge, which restrict Ag to the $1+$ oxidation state \cite{SM}. Moreover, we find a six orders of magnitude reduction in low temperature resistivity of \AgIr{} compared with \Li{} \cite{SM}. Thus, the more plausible explanation is that the increased electronegativity of Ag over Li results in an larger inductive effect on Ir, bringing Ir away from half-filling. This is consistent with an increase in Ir-O hybridization in \AgIr{}  when compared to \Li{} as found by LDA+$U$ calculations \cite{PhysRevLett.123.017201}.

%---------------------------------------
\begin{table}[!t]
    \centering
    \begin{tabular}{c c c c c}
       Material ($\langle n_h \rangle$)& L$_3$ (eV) & I(L$_3$)+ I(L$_2$) & BR & $\LS $ \\
      \hline
      \hline
      {\Li} (5) & 11219.0 & 22.0(5) & 5.0(4) & 2.5(2)\\
      {\AgIr} (5.5) & 11219.4 & 26.0(5) & 5.7(4) & 3.1(2)\\
      Sr$_3$CaIr$_2$O$_9$ (6) & 11220.0 & 30.0(5) & 6.5(5) & 3.6(2)
    \end{tabular}
    \caption{Summary of the parameters extracted from the analysis of the L$_{3,2}$ XAS data.}
    \label{tab:XAS}
\end{table}

Metal ligand hybridization is known to affect the magnetism in transition-metal oxides \cite{Mazurenko2015, Streltsov10491,Streltsov_2017,PhysRevB.93.214431}. We performed XMCD at the Ir L$_{3,2}$-edges with $H$=5~T and $T$=1.6 K in order to understand the influence of the charge redistribution on the magnetism [Fig.~\ref{fig:Absorption} (c) -- (d)]. XMCD measures the projection of the magnetic moment along the x-ray helicity, set to be parallel to the magnetic field direction. We find that the XMCD signal at the $L_3$ edge is reduced in \AgIr{} compared with \Li{}. This may be related to a reduction in the on-site Ir moment, or to differences in the in-field magnetic structure between these two compounds. We apply XMCD sum rules to find orbital $\Lz$ spin $\Sz$, and intra-atomic magnetic dipole moment $T_z$ contributions \cite{PhysRevLett.68.1943, PhysRevLett.70.694, PhysRevLett.75.152}. For a single $j_{\rm eff} \!=\! 1/2$ hole one expects $\Lz/2\Sz \!=\! 2$  \cite{PhysRevLett.101.076402}. Small departures from the ideal values arising from non-cubic crystal fields and covalency are common in iridates \cite{ PhysRevLett.109.027204, PhysRevB.90.014419, PhysRevB.97.214436}, but we find even larger differences for \AgIr{}. Sum rule analysis using the measured $\mu(H = 5 \rm{T}) \approx 0.05~\mu_B/Ir$ moment for \Li{} \cite{PhysRevB.99.054426} gives $\Lz \!=\! -0.028(1)~\mu_B/Ir$, $\Sz \!=\! -0.011~\ub$ and $\Lz/2\Sz\!=\! 1.27$ and $7\Tz/2\Sz \!=\! 1$. For \AgIr{} we find $\Lz \!=\! -0.018(3)~\mu_B/Ir$ and $\Sz \!=\! -0.002(3)~\ub$, using $\langle n_h \rangle = 5.5(2)$ and the measured moment of $\mu(H = 5 \rm{T})\!=\! (\Lz+2\Sz) \!=\! 0.022(5)~\ub$ \cite{SM,bahrami2020effect}. These values lead to $\Lz/2\Sz \!=\!2.6-4.5$ and $7\Tz/2\Sz \!=\! 1.85 - 6.15$.  The minimum twofold enhancement of the fractional contribution of $\Lz$ to the total moment in \AgIr{} further rules out Ir$^{5+}$  $j = 0$ states where $\Lz/2\Sz \!=\! 0.5$ and establishes the orbital nature of magnetism. 

Large values of $T_z$ are associated with aspherical spin density originating from non-cubic fields, spin-orbit coupling, and electron delocalization due to metal-ligand hybridization \cite{STOHR1999470, PhysRevLett.88.207203, Schmitz2014}. To quantify departures from a cubic Ir environment, we carefully examined the local structures of \AgIr{} using x-ray PDF and EXAFS measurements  [Fig.~\ref{fig:structure}]. Both measurements are independently consistent with the reported $C2/m$ space group \cite{PhysRevLett.123.237203} over a broad $r$-range. We find distortions of the Ir environment, with Ir-O distances ranging from 2.021 to 2.119~\AA{}, and  Ir-O-Ir bond angles of 92.38$^{\circ}$ and 95.72$^{\circ}$ \cite{SM}. While the associated trigonal fields in \AgIr{} are larger than in \Li{}, the $5\%$ difference in Ir-O bond distance is not enough to account for the minimum $80\%$ enhancement of $7\Tz/2\Sz$.  We thus assign differences in the magnetism of \AgIr{} to an increased delocalization of the Ir $5d$ electrons over the ligands \cite{PhysRevLett.88.207203}.

\begin{figure}[!t]
  \begin{center}
    \includegraphics{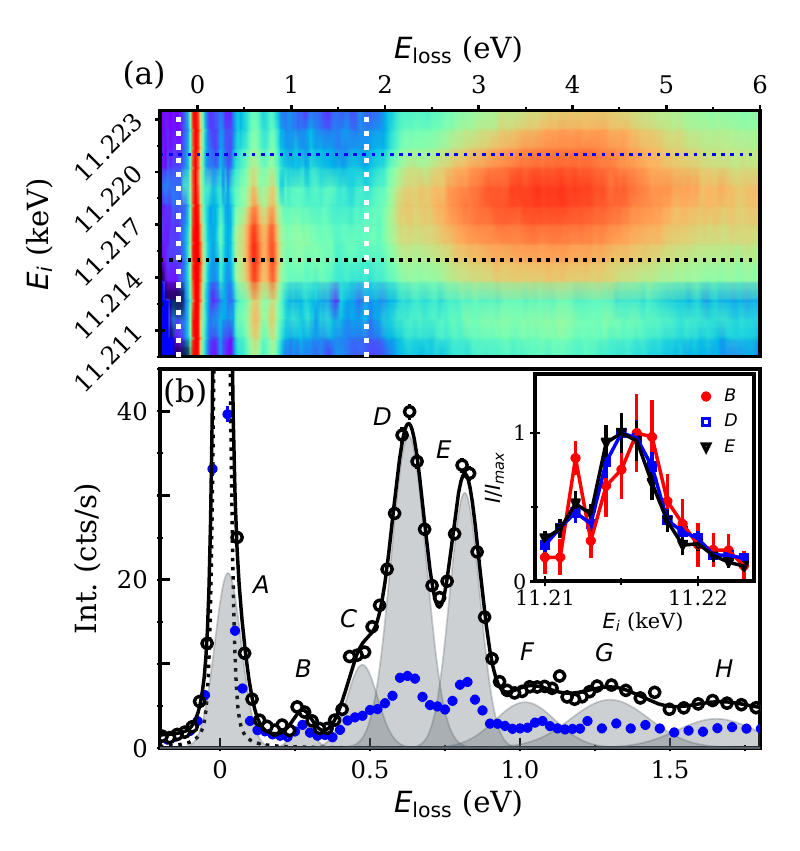}
  \caption{(a) Ir L$_3$ RIXS intensity as a function of $E_i$ and $E_\mathrm{loss}$ in \AgIr{} at $T \!= \!20$~K. Vertical dotted lines indicate the\eloss{} range for (b). Horizontal dotted lines are color codded according to the RIXS spectrum in (b), (b) Representative RIXS spectrum at $E_i = 11.215$ keV (black markers) and fit to the data (black solid line) with Gaussian peaks (grey shaded) and Voigt elastic line (dotted), as described in the main text. Blue markers show the RIXS spectrum  at $E_i = 11.221$ keV. Inset shows the integrated intensity of features $B$, $D$ and $E$ as a function of $E_i$}.
  \label{fig:Overview}
  \end{center}
\end{figure}

%--------Figure 4 -------------
\begin{figure*}[!ht]
  \begin{center}
    \includegraphics[width=1\textwidth,clip]{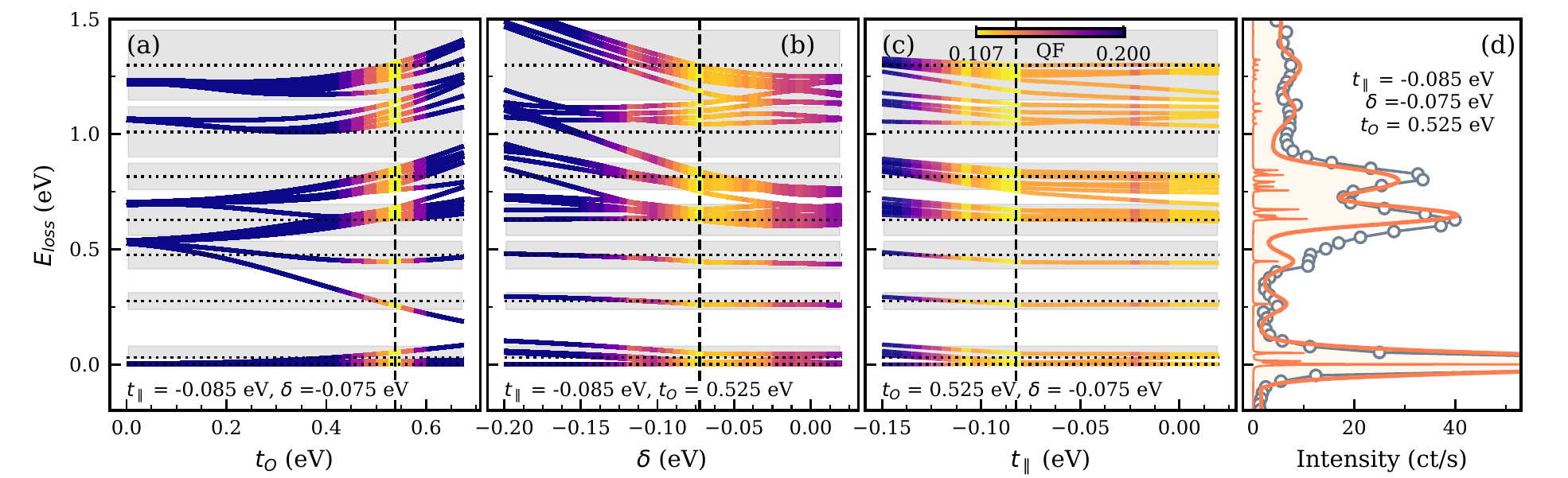}
  \caption{Calculated energy spectrum of \AgIr{} as a function of (a) $t_O$, (b) $\delta$ and (c) $t_{\parallel}$ with $\lambda\! =\! 0.395 $ eV, $U\! =\! 2$ eV, $J_H\! =\! 0.3$ eV, $t_{\sigma}\! =\! - t_{\parallel}$, and $t_{\perp}\! =\!-0.05~t_{\parallel}$. Horizontal dashed line and grey boxes represent the center energy and FWHM of the extracted inelastic peaks. The color map indicates agreement of the calculation with the experimental data as encoded by the quality factor (QF). (d) Simulated RIXS spectrum for the best parameters compared to the measured intensity.}
  \label{fig:Model}
  \end{center}
\end{figure*}

Having found that the magnetism in \AgIr{} is characterized by a more delocalized electron density, we performed RIXS measurements to better understand the Ir electronic structure. Fig.~\ref{fig:Overview} (a), shows the $T\! =\! 20$~K RIXS signal in {\AgIr} as a function of incident energy, $E_{\rm i}$ and energy transfer, \eloss{}. Thompson scattering has a vanishing cross-section in our 90 degree scattering configuration, thus any phonon contribution to the RIXS signal is minimal. The broad and intense features centered at {\eloss} $ \approx 3.8$ eV, originate from $t_{2g} \rightarrow e_g$ transitions, consistent with the strong crystal field limit, $10Dq > 3$ eV \cite{PhysRevLett.109.157401,PhysRevLett.110.076402,PhysRevLett.122.106401,Clancy2018}. For the remainder of this work, we concentrate on the intra-t$_{2g}$ excitations, \eloss{}$\leq1.5$~eV. Fig.~\ref{fig:Overview} (b) shows a representative RIXS spectrum of \AgIr{} at $E_i\!=\!11.215$~keV. In order to analyze the experimental data, we fit eight Gaussian peaks: $A\!=\!0.029(25)$, $B\!=\!0.270(37)$, $C\!=\!0.470(60)$, $D\!=\!0.623(69)$, $E\!=\! 0.811(58)$, $F\!=\!1.010(109)$, $G\!=\!1.300(153)$, $H\!=\! 1.654(150)$ eV and include a resolution limited Voigt function for the elastic line \cite{SM}. The RIXS spectrum of \AgIr{} is distinct from an ideal local $j_{\rm eff} \! = \! 1/2$ state, where only a single excitation between the $j \! = \! 3/2$ quartet and $1/2$ doublet appears at  $\Delta E\!=\!3\lambda/2 \! \approx\! 0. 75$~eV, where $\lambda$ is the spin orbit coupling constant. $\Delta E_{D-E}\!=\!0.19(4)$~eV is larger than the corresponding 0.11(3)~eV splittings in \Li{} and Na$_2$IrO$_3$ \cite{PhysRevLett.110.076402}, consistent with a more trigonally distorted environment. However, non-cubic crystal fields (CF) cannot reconcile the RIXS spectra with a single-site Ir$^{4+}$ model that can at most produce two RIXS features for $E_{\rm loss}\! <\! 1$~eV \cite{SM}.

The energy of $A$, $E^A_{\mathrm{loss}}= 29$~meV, is consistent with collective magnetic excitations observed in other iridates \cite{PhysRevX.10.021034,PhysRevResearch.2.043094,PhysRevB.87.220407, ruiz2021magnonspinon, PhysRevB.103.L020410}. We find that the intensity of peak $A$ follows detailed balance; but there is no correlation with the Curie-Weiss temperature ($\theta_{CW}\!=\!-132$~K) \cite{bahrami2020effect}, ruling out a spin wave origin \cite{SM}.  Peak $C$, $E^C_{\mathrm{loss}}= 470$~meV, is suggestive of the electron-hole exciton in Na$_2$IrO$_3$, \Li{}, and Sr$_2$IrO$_4$ \cite{PhysRevLett.110.076402, Kim2014}. Peak $B$, $E^B_{\mathrm{loss}} = 0.279 $ eV, is more puzzling. A similar feature in other Ir$^{4+}$ compounds was associated with Ir$^{5+}$ impurities based on qualitative arguments, but the relative intensity of this feature rules out an origin from the $<1\%$ off-stoichiometries in our sample, as constrained by chemical analysis \cite{SM, bahrami2020effect}.

Our XAS and XMCD measurements indicate strong Ir-O hybridization  in \AgIr{}. Thus, we consider a quasi-local model including trigonal fields and hopping integrals between nearest-neighbor Ir atoms in the large cubic crystal field limit ($t_{\parallel}$, $t_{\rm O}$, $t_{\sigma}$, $t_{\perp}$). We follow the labels and hopping paths described in Ref.~\cite{PhysRevB.88.035107}. These hopping terms are akin to those discussed for the QMO state  \cite{PhysRevLett.109.197201,PhysRevB.88.035107,PhysRevMaterials.4.075002,Streltsov_2017}. We note that our model nearest-neighbour model is the minimal unit for electron delocalization ant is realistic approximation that captures hybridization related splitting of single ion levels \cite{PhysRevLett.122.106401}. We fix $\lambda \!=\! 0.395 $ eV, on-site Coulomb interaction, $U \!=\! 2$ eV, and Hund's coupling, $J_H \!=\! 0.3$ eV, \cite{SM,PhysRevLett.122.106401}. In, Fig.~\ref{fig:Model} (a) -- (c), we show the resulting energy spectrum from exact diagonalization calculations as a function of the dominating $t_{2g}$ hopping integrals $t_{\parallel}$, hopping between orbitals in parallel plane, and $t_O$ , O $2_p$ mediated hopping, and non-cubic crystal fields, $\delta$. The dashed horizontal lines and shaded boxes indicate the centroids and FWHM of peaks $A$ -- $G$. We parameterize the energy difference between the calculated excitonic spectrum and the experimental peak positions through a \textit{quality factor}, QF \cite{SM} minimized across $\delta$, $t_{\parallel}$, $t_{\rm O}$, $t_{\sigma}\!=\!-t_{\parallel}$, and $t_{\perp} \!=\! -0.05 t_{\parallel}$. Measured RIXS spectra can be reproduced within our model for the parameter set $t_{\rm O} \!=\! 0.525(13)$~eV, $ -0.18 \! < \! t_{\parallel} \!<\! -0.07$~eV, and $ - 0.07 \! < \! \delta \! < \! -0.02$~eV with a 10\% variation of QF from the minimum \cite{SM}. Density functional theory calculations find $t_{\parallel}/t_{\rm O} \approx 0.1$ in H$_3$LiIr$_2$O$_6$ \cite{PhysRevLett.121.247202}; thus, we fix $t_{\parallel}$ to the lower end of the range consistent with our data. False colormaps in Fig.~\ref{fig:Model} (a) -- (c) show the dependence of the QF around a set of parameters within this range. Using the optimized parameters: $t_{\rm O} \!=\! 0.525$~eV, $t_{\parallel} \!=\! -0.085$~eV, and $\delta \!=\! -0.075$~ eV, we computed the powder averaged RIXS intensity over the three inequivalent bond directions \cite{SM, PhysRevLett.121.247202,WANG2019151} and convolved with a Gaussian profile (FWHM = 33~meV) to compare with our data in Fig. ~\ref{fig:Model} (d).  

 We can account for all observed \eloss{} $\!<\! 1.5$~eV RIXS peaks, including peaks $A$, $B$, $C$, $F$ and $G$ which the single ion model completely fails to capture by including electron hopping and non-cubic crystal fields. In particular, peak $B$ is strongly dependent on $t_{\rm O}$. This hopping term is inversely proportional to the energy difference between Ir $t_{2g}$ and O $2p$ orbitals and \textit{ab-initio} calculations have shown that it depends on the intercalated ion. In H$_3$LiIr$_2$O$_6$, hybridization of the H $1s$ and O $2p$ orbitals enhances $t_{\rm O} \!=\! 0.3 - 0.4$~eV \cite{PhysRevLett.121.247202} with respect to that of Na$_2$IrO$_3$ ($t_{\rm O}\! =\! 0.270$~ eV) \cite{PhysRevLett.109.197201, PhysRevB.88.035107}. It is plausible that the larger spatial extent of Ag orbitals compared with Li, and the energy overlap of Ag and O states \cite{PhysRevLett.123.237203} leads to the larger $t_{\rm O}$ in \AgIr{}. Including further neighbor terms may also reduce the value of $t_{\rm O}$ needed to explain our data \cite{PhysRevB.93.214431}. Our values of $\lambda/t_{\rm O} = 0.75$ and $(U-3J_H)/t_{\rm O} \!=\! 2.66$ set \AgIr{} well within the delocalized regime \cite{PhysRevLett.117.187201}. Similar effects could be at play in H$_3$LiIr$_2$O$_6$ and we find that hyrbridization-related RIXS features should be resolvable for values of $t_{\rm O} \!>\! 0.2$ eV, within expectations for that compound. 

The delocalized magnetic state in \AgIr{} has important implications for the magnetism of intercalated iridates. In \AgIr{}, the paramagnetic moment $\mu_{\mathrm{eff}}\! =\! 1.87~\mu_B$/Ir \cite{PhysRevLett.123.237203,bahrami2020effect} agrees with the expected 1.74~$\mu_B$/Ir for $j_{\rm eff}\! =\! 1/2$. Muon spin resonance measurements also find oscillations characteristic of incommensurate magnetic order, similar to that of \Li{}, hinting at a common origin of the magnetism \cite{bahrami2020effect,PhysRevB.93.214431}. Our finding of an asymetric spin density and large orbital contribution  warrants a more nuanced interpretation. The fundamental magnetic unit in \AgIr{} is an extended and anisotropic electron density over Ir and O sites \cite{li2021modified} that will promote long-range and possibly anisotropic interactions. The result is a greater magnetic frustration with enhanced Curie-Weiss, but similar ordering temperatures to the parent \Li{} \cite{PhysRevLett.123.237203, bahrami2020effect}. Thus, itineracy could offer another possible route to spin liquids in iridates. The spatially extended spin density also has consequences for neutron scattering measurements, as the magnetic form-factor of such a state will depart significantly from the localized limit. In order to interpret magnetic measurements, more detailed studies of the effective magnetic Hamiltonians for such a delocalized state are needed \cite{PhysRevLett.121.247202}.

In summary, we have used a suite of x-ray spectroscopies to find that intercalation of Ag atoms on inter honeycomb layer sites in \Li{} promotes Ir-O hybridization and alters the magnetism. Our data and analysis suggest that the electronic structure \AgIr{} is nearly itinerant, a result of a more trigonally distorted Ir environment and enhanced hopping integrals. The magnetism is characterized by a delocalized asymmetric spin density and large orbital moment that is drastically different from the parent compound. Magnetic frustration is enhanced indicating itineracy may offer an alternative route to spin liquids in the iridates. This phenomenology may extend to other intercalated versions of the honeycomb iridates and our results provide a foundation to develop effective magnetic Hamiltonians for these compounds. 

We thank Mark Dean for helpful discussions and for his critical reading of this manuscript. Work at Brown University was supported by the U.S. Department of Energy, Office of Science, Office of Basic Energy Sciences, under Award Number DE-SC002165. The work at Boston College was supported by the National Science Foundation under award No. DMR– 1708929. TMM and JC acknowledge support from the Institute for Quantum Matter, an Energy Frontier Research Center funded by the U.S. Department of Energy, Office of Science, Office of Basic Energy Sciences, under Award DE-SC0019331. Use of the Advanced Photon Source at Argonne National Laboratory was supported by the U. S. Department of Energy, Office of Science, Office of Basic Energy Sciences, under Contract No. DE-AC02-06CH11357.

\pagebreak
\clearpage
\begin{widetext}
\begin{center}
\textbf{\large Supplemental Material For: Nearly itinerant electronic groundstate in the intercalated honeycomb iridate \AgIr{}}
\end{center}
\end{widetext}

\renewcommand{\thefigure}{S\arabic{figure}}
\renewcommand{\thetable}{S\arabic{table}}
\renewcommand{\theequation}{S\arabic{equation}}
\setcounter{figure}{0}
Here we provide additional technical details describing: \begin{enumerate*}[label=(\arabic*)]
    \item \AgIr{} chemical characterization.
    \item Temperature dependent resistivity in \AgIr{} and \Li{}. 
    \item Analysis and fitting of x-ray absorption data.
    \item Temperature dependence of XAS spectra.
    \item XAS at the Ag L$_3$-edge in \AgIr{}.
    \item Additional RIXS data for Ir$^{4+}$ and Ir$^{5+}$ compounds.
    \item Temperature dependence of the RIXS spectra. 
    \item Details of the EXAFS refinements. 
    \item Comparison of bulk magnetisation and XMCD signal as a function of field.
    \item Structural parameters of \AgIr{} refined from pair distribution and EXAFS measurements.
    \item The crystal field Hamiltonian used for treatment of crystals field excitations in the RIXS spectra.
    \item Details of Slater integrals used for single site calculation.
    \item The tight bonding Hamiltonian used to model the RIXS spectra including electron hopping.
    \item Role of Hund's coupling, $J_H$, and electron-electron correlations, $U$, in the calculated RIXS spectrum.
    \item Details of the \textit{quality factor} used to optimize parameters of our tight binding model against the measured RIXS spectra. 
    \item Procedure for computing the powder (spherically) averaged RIXS spectra. 
\end{enumerate*} 
\section{{A\lowercase{g}$_3$L\lowercase{i}I\lowercase{r}$_2$O$_6$}  chemical characterization}

\begin{figure}[!ht]
  \begin{center}
    \includegraphics[width=0.4\textwidth,clip]{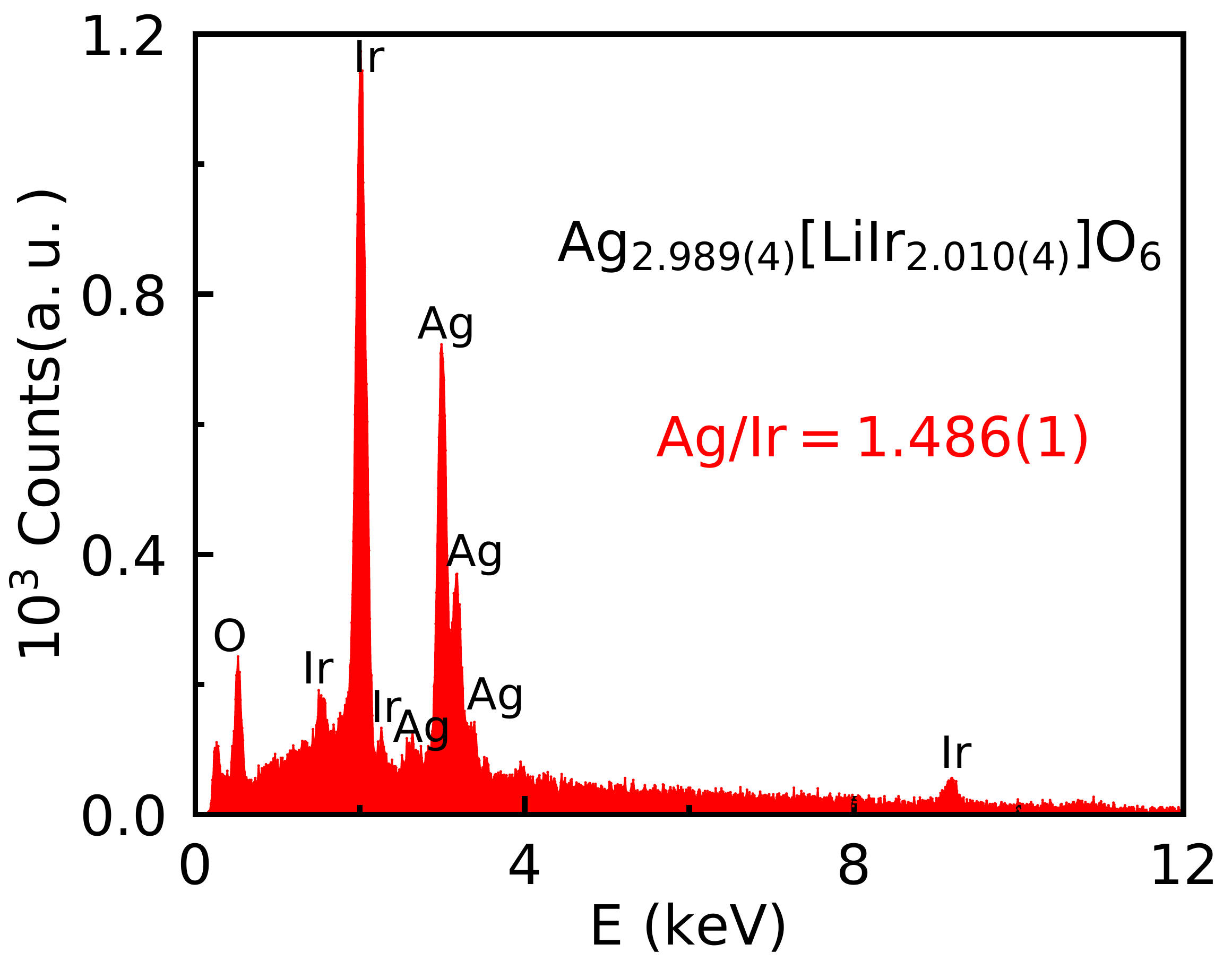}
  \caption{Energy-dispersive x-ray spectroscopy (EDX) spectrum  from \AgIr{}}
  \label{fig:EDX}
  \end{center}
\end{figure}

In Fig.~\ref{fig:EDX}, we show the energy-dispersive x-ray spectroscopy (EDX) spectrum averaged from three different spots on a pellet sample of \AgIr{} from the same batch of sample used for our RIXS and XAS measurements. This is also the same high quality sample used in \cite{SI_bahrami2020effect}. The mol ratio between Ag and Ir is $\rm{Ag}/\rm{Ir} = 1.486(1)$, as expected from the chemical formula. Although a more precise characterization of this ratio is hindered by the sensitive of the EDX technique, our measurements indicate that to keep charge balance the maximum amount of Ir$^5+$ is limited to $0.2\%$. This is also consistent with the extracted paramagnetic moment of $\mu = 1.87(2) \mu_B$ from magnetic susceptibility, \cite{SI_bahrami2020effect}. The presence of off-stochiometric Ir $^{5+}$, would result in a reduced paramagnetic magnetic moment. Within the error bars of our magnetization measurements and assuming $g \approx 2.4$ to account for the large moment when compared to that of a pure $J_{eff} = 1/2$ groundstates \cite{SI_li2021modified}, we can rule out Ir$^{5+} > 1\%$.

\section{Resistivity comparison}

\begin{figure}[!ht]
  \begin{center}
    \includegraphics[width=0.4\textwidth,clip]{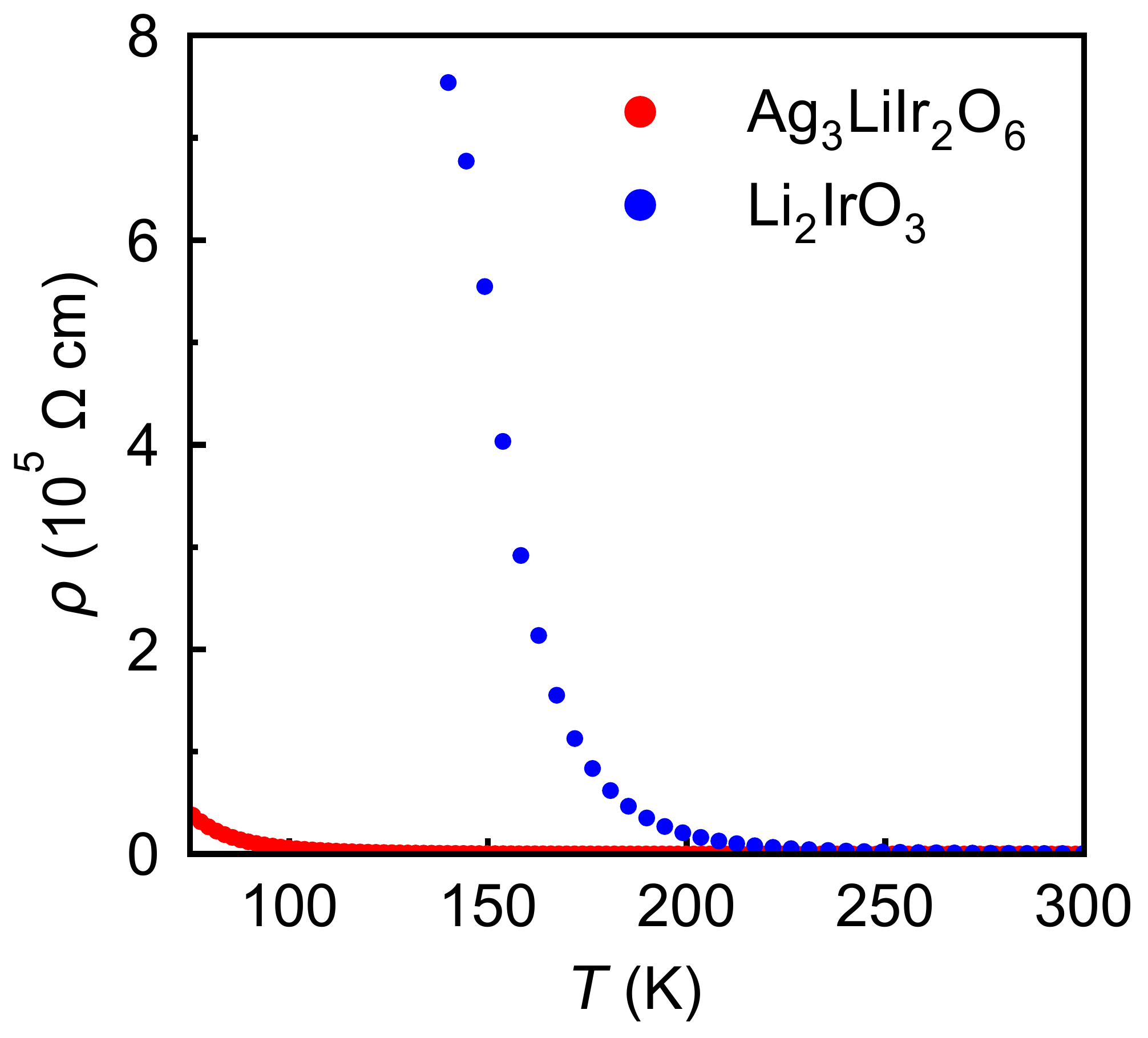}
  \caption{Comparison between the resistivity of \AgIr{} (red) and \Li{} (blue)}
  \label{fig:resistivity}
  \end{center}
  
\end{figure}

Fig.~\ref{fig:resistivity} shows the measured resistivity of pelletized powder samples of \AgIr{} and \Li{}. Both compounds are insulating, with room temperature resistivities of the order Ohm $\cdot$ cm. However, below 200~K, the resistivity of \AgIr{} becomes six orders of magnitude smaller than \Li{}. The significantly reduced resistivity of \AgIr{} strongly supports the conclusion of a more covalent Ir state, as  observed by XAS and RIXS, with a larger Ag-mediated hybridization between Ir and O orbitals.

\section{Details of the XAS analysis}
\begin{figure}[!ht]
  \begin{center}
    \includegraphics{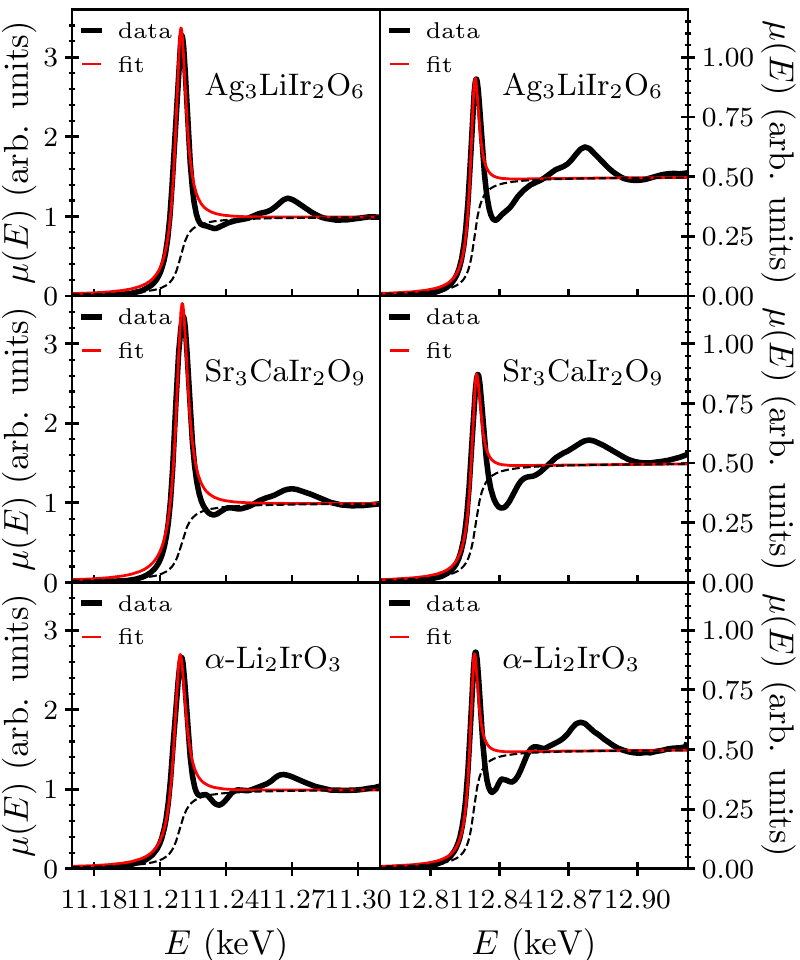}
  \caption{XAS at the Ir L$_3$ and L$_2$ absorption edges at $T$=300~K. The red line is a fit to the experimental data (black lines). Note the different y-axis scales between the Ir L2 and L3 edge data.}
  \label{fig:XAS}
  \end{center}
\end{figure}
XAS data was analyzed using the Larch software package \cite{SI_Newville_2013}. Data was normalized using standard protocols by first fitting both the \textit{pre-edge} region and the \textit{post-edge} region to a linear term. The step height was found as the difference between the two fitted lines at the edge position. The edge position $E_0$ was found from the inflection point of the data. Following established procedures, the pre-edge fit was subtracted from the data, which was then normalized to step heights of 1 and $\frac{1}{2}$ for the $L_3$ and $L_2$ edges respectively. Finally, a quadratic fit to the post-edge region was subtracted to obtain the fully normalized data. The normalized data was subsequently fitted to an arctangent step and Lorentzian peak to extract the parameters of the white line. This is shown in Fig.~\ref{fig:XAS} the normalized XAS data for \AgIr{}, \Li{} ($4+$) and Sr$_3$CaIr$_2$O$_9$ ($5+$). 

\begin{figure}[!ht]
  \begin{center}
    \includegraphics{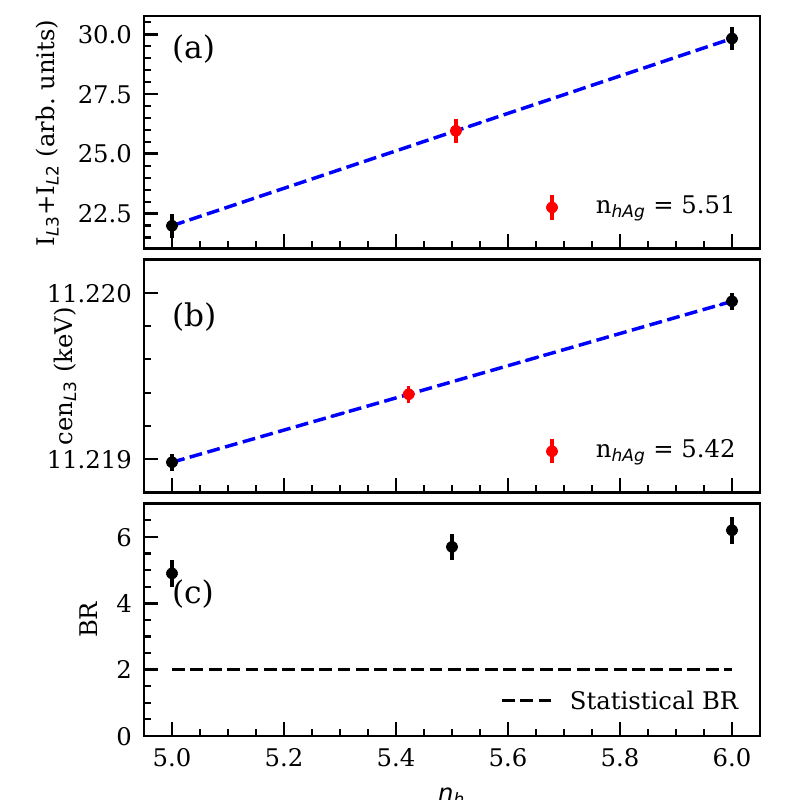}
  \caption{Number of holes extracted from the XAS data from (a) sum rules and (b) L$_3$ inflection point. (c) Branching ratio extracted from L2 and L3 white line intensities.}
  \label{fig:xaspar}
  \end{center}
\end{figure}
The number of holes, $\langle n_h \rangle$, in the $5d$ orbitals can be calculated from the XAS sum rule: $I_{L3} + I_{L2} = \langle n_h \rangle C$, with $C$ a proportionality constant \cite{SI_Klemradt:668996}. Additionally, a monotonic trend of the position of the inflection point of the $L_3$ absorption edge towards higher energies with increasing number of holes has been observed in iridates \cite{SI_PhysRevB.86.195131}. By fitting the data to a $y = y_0 + Ax$ we find for the sum rule  $\langle n_h \rangle = 5.5(2)$ while $\langle n_h \rangle = 5.4(2)$ for the position of the inflection points. We calculated $\langle n_h \rangle = 5.5(2)$ for \AgIr{} by averaging the results using these two approaches in Fig.~\ref{fig:xaspar} (a) and (b), with \Li{} $\langle n_h \rangle = 5$ and Sr$_3$CaIr$_2$O$_9$ $\langle n_h \rangle = 6$. 

The branching ratio $BR\! =\! I_{L_3}/I_{L_2}$ extracted from numerical integration and calculated from the fit parameters is shown in Fig.~\ref{fig:xaspar} (c) compared to the statistical value (black dashed line). The BR remains larger than the statistical BR signifying the importance of spin-orbit coupling for the three compounds.

\section{XAS temperature dependence}
\begin{figure}[!ht]
  \begin{center}
    \includegraphics{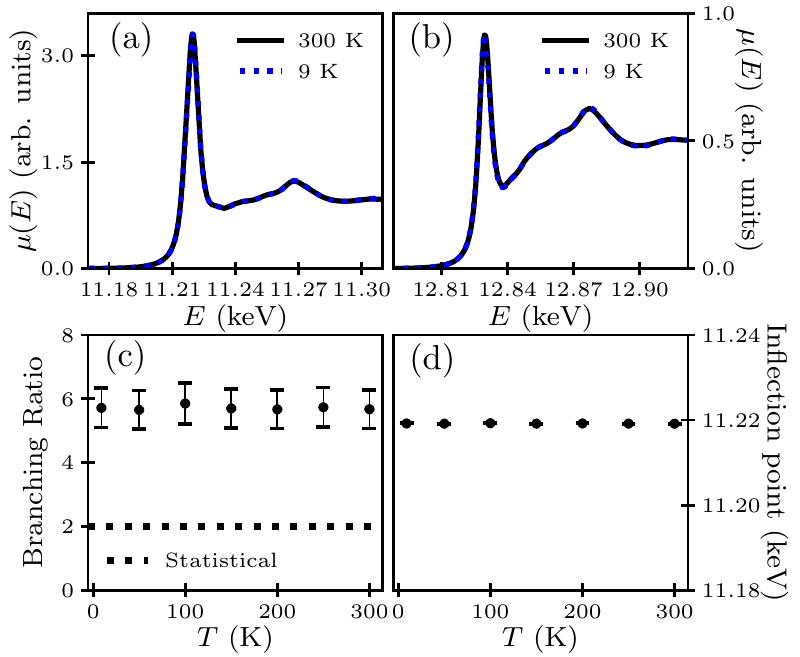}
  \caption{(a),(b) Representative normalized XAS data of \AgIr{} at the L$_{3,2}$ at $T \!=\! 300$~K and $T \!=\! 9$~K. Temperature dependence of the (c) BR and (d) inflation point of the $L_3$  }
  \label{fig:xasT}
  \end{center}
\end{figure}
In Fig.~\ref{fig:xasT}, we show the \AgIr{} XAS data at $T \!=\! 300$~K and $T \!=\! 9$~K at the L$_3$, (a), and L$_2$, (b), absorption edges. We observe no temperature dependence of the BR, (c), and , (d), inflection point of the L$_3$ white line. 

\section{A\lowercase{g} L$_3$ edge XAS}

\begin{figure}[!ht]
  \begin{center}
    \includegraphics{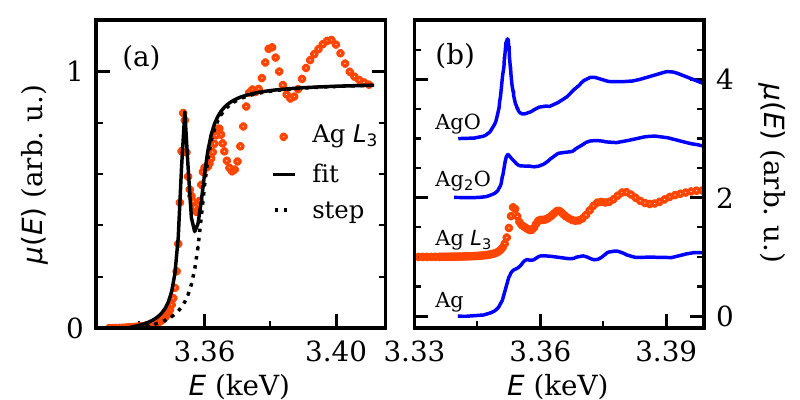}
  \caption{(a)Representative normalized XAS data of \AgIr{} at the Ag L$_{3}$ at $T \!=\! 300$~K, (b) compared to that of Ag ($0^+$), Ag$_2$O ($1^+$) and AgO ($2^+$) adapted from \cite{SI_Kolobov_ag}. Spectra were vertically shifted for clarity and horizontal to match the white line position.}
  \label{fig:xasAg}
  \end{center}
\end{figure}

In Fig.~\ref{fig:xasAg} (a), we show  $T \!=\! 300$~K XAS data collected at the Ag $L_{3}$ edge of our \AgIr{} sample. Data were collected in a partial yield  fluorescence geometry using a four-element Si-drift diode. The intensity of the white line peak  can be related to the number of holes in the Ag $4d$ orbitals and the degree of covalence of the Ag bonds \cite{SI_behrens_Ag}. The data shows a characteristic absorption peak for Ag $1+$ at $\rm{E_{i}} =3.349$ keV, with comparable intensity to the post edge features.  The XAS spectrum and the intensity of this absorption feature in \AgIr{}, $I_{L_3} = 0.87(3)$, resemble those of Ag$_2$O \cite{SI_Kolobov_ag}, Fig.~\ref{fig:xasAg} (b) and we conclude from this data that Ag is in the $1+$ oxidation state in \AgIr{}, further ruling out Ir$^{5+}$ impurities.

\section{Comparison of RIXS signal in I\lowercase{r}$^{4+}$ and I\lowercase{r}$^{5+}$ compounds}
\label{comp_rixs}

As discussed in the main text, an inelastic feature near $300$ meV has previously been observed and  assigned to Ir$^{5+}$ impurities in a highly distorted Ir $^{4+}$ double perovskites Sr$_2$CeIrO$_6$ \cite{SI_PhysRevB.99.134417}; Sr$_3$NiIrO$_6$, a compound with two magnetic species \cite{SI_PhysRevB.93.224401}; in an iridium fluoride with two nonequivalent Ir sites,  Na$_2$IrFl$_6$ \cite{SI_PhysRevB.95.235161}; and in the dimer material  Ba$_5$AlIr$_2$O$_{11}$ \cite{SI_PhysRevLett.122.106401}. We note that in each case mentioned above this assignment was based on qualitative comparison and that no previous quantification of this signal has been reported. We now rule out the possibility of an Ir$^{5+}$ RIXS signal in \AgIr{} based on a direct comparison of the RIXS cross-section with a known Ir$^{5+}$ standard.

\begin{figure}[!ht]
  \begin{center}
    \includegraphics{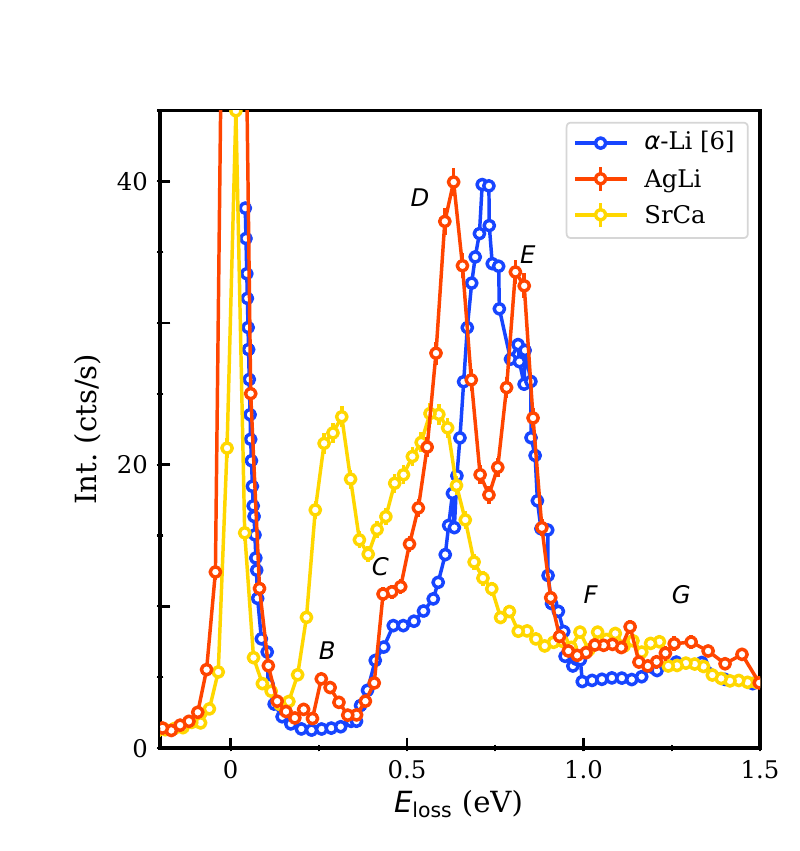}
  \caption{Direct comparison of the RIXS spectrum at resonance,$E = 11.215$ keV, between 
  \AgIr{} (orange), Sr$_3$CaIr$_2$O$_9$ (yellow) measured under the same experimental circumstances, and \Li{} (blue) adapted from \cite{SI_PhysRevLett.110.076402}. The data for \Li{} has been scaled by an arbitrary factor for qualitative comparison.}
  \label{fig:4p5p} 
  \end{center}
\end{figure}

In Fig.~\ref{fig:4p5p}, we show RIXS spectrum at  $\rm{E_{in}} = 11.215$ keV for both \AgIr{} and Sr$_3$CaIr$_2$O$_9$, a model Ir$^{5+}$ compound. The measurements were collected consecutively under identical experimental conditions and therefor provide a direct quantification of the relative contributions of Ir$^{4+}$ and Ir$^{5+}$ to the RIXS cross-section. As expected for Ir $d^4$ in an IrO$_6$ octahedral environment, the low energy RIXS spectrum of Sr$_3$CaIr$_2$O$_9$ is characterized by an inelastic feature at \eloss{} $=330(25) $ meV and a broader peak at \eloss{} $=600(50)$ meV, similar to other Ir $5^+$ compound\cite{SI_PhysRevLett.123.017201,SI_PhysRevB.95.235114}. We note that these RIXS features in Ir$^{5+}$ compounds are not sensitive to distortions of the IrO$_6$ octahedra \cite{SI_PhysRevB.95.235114}.

The scattering volume is the same between samples and the total cross-section scales with the number of scattering nuclei in the unit cell volume $N = n/V$.  In  Sr$_3$CaIr$_2$O$_9$, $N = 0.0101$ and in  \AgIr{} $N = 0.013$. Thus, we can use the integrated intensity of the 330 eV peak in Sr$_3$CaIr$_2$O$_9$ to estimate the molar percentage of Ir $5^+$ required to produce a corresponding RIXS signal in \AgIr{} $I(B)$. We find  that  $I^{300}_{\rm{SrCa}} = 24(1) $ cts/s and correspondingly  $\approx 20\%$ Ir $^{5+}$ in \AgIr{} would be required to produce the measured intensity of $I(B) = 4.8(2)$ cts/s. This is two orders of magnitude greater than the Ir${^5+}$ concentration allowed by chemical characterization and magnetic susceptibility measurements. Thus, while we cannot rule out Ir$^{5+}$ impurities at the level of 1\% or less, such a small impurity concentration will only have a negligible contribution and cannot account for the measured RIXS signal.

Additionally, in Fig.~\ref{fig:4p5p} we also reproduce \Li{} RIXS, adapted from \cite{SI_PhysRevLett.110.076402}. We remark that the difference in the low energy features with respect to \AgIr{} are not restricted to the large splitting between peaks $D$ and $E$ and a clear peak $B$, but also to much more defined $F$ and $G$ features.

\section{RIXS temperature dependence}
\begin{figure}[!ht]
  \begin{center}
    \includegraphics{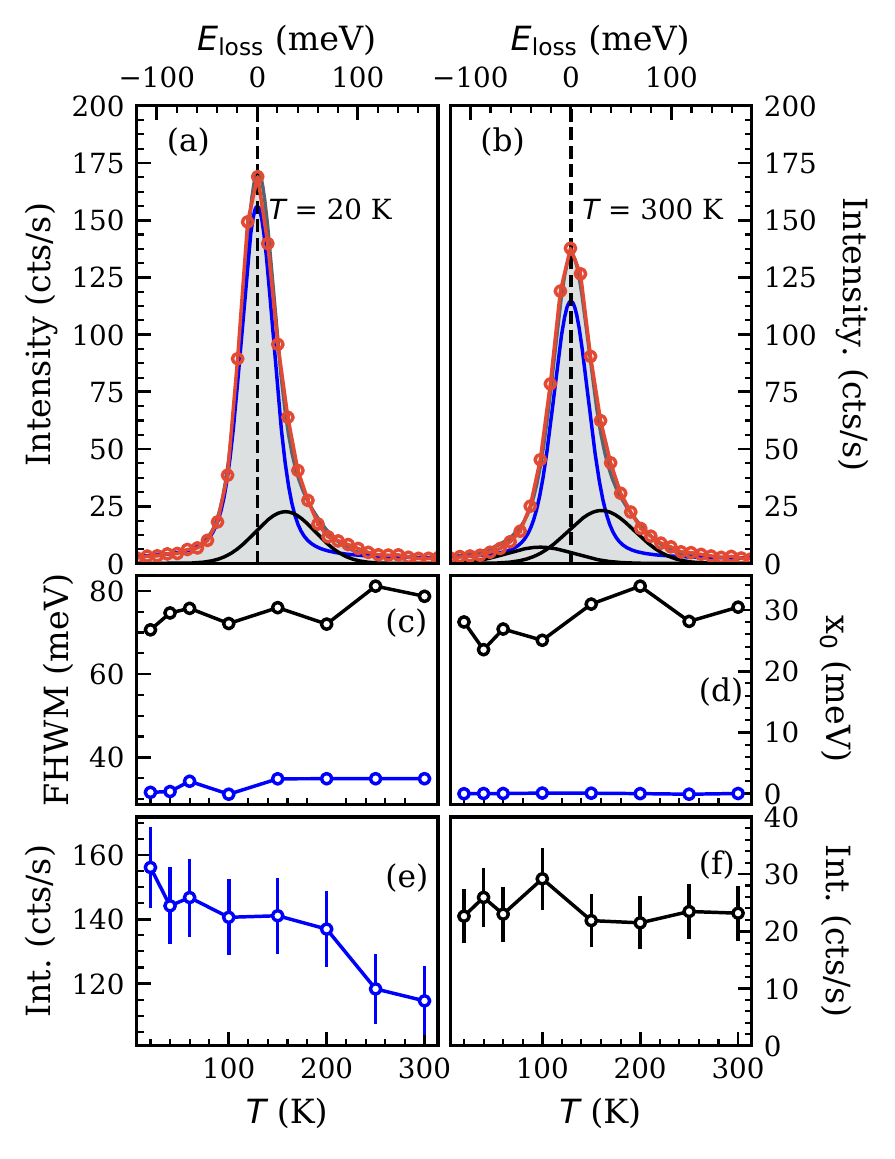}
  \caption{(a)-(b) RIXS data (red markers) near the elastic line and fit to the data (grey shading). Blue line indicates the resolution limited Voigt line (blue line) and a Gaussian peak for feature $A$ including detailed balance. Temperature dependence of (c) FWHM, (d) centroid and (e)-(f) intensity. Color coded markers indicate elastic line and Peak A.}
  \label{fig:eline}
  \end{center}
\end{figure}

In Fig.~\ref{fig:eline} (a) -- (b) we show a magnified region of the RIXS data near the elastic line for $T\!=\! 20$~K and $T\!=\! 300$~K. Within the energy range \eloss{} $\in [-100,100]$~meV we fit the data to a resolution limited Voigt line (blue line) and a Gaussian peak for feature $A$ constrained by detailed balance (black line). No temperature dependence of the FWHM (c), centroid (d), or intensity (f) is observed for Peak $A$.

We observe an increase of intensity in the elastic line with decreasing temperature. At $2\theta = 90^{\circ}$ we are in the proximity of structural Bragg peaks, for example $[2,8,-6]$ at $2\theta = 89.77^{\circ}$ at $\lambda = 1.1055$ \AA{}. Thus, it is likely that this increased elastic intensity is a result of lattice parameters contractions which bring the tail of the Bragg peak into the measurement window, but a weak magnetic signal is also a possibility.  

\begin{figure}[!ht]
  \begin{center}
    \includegraphics{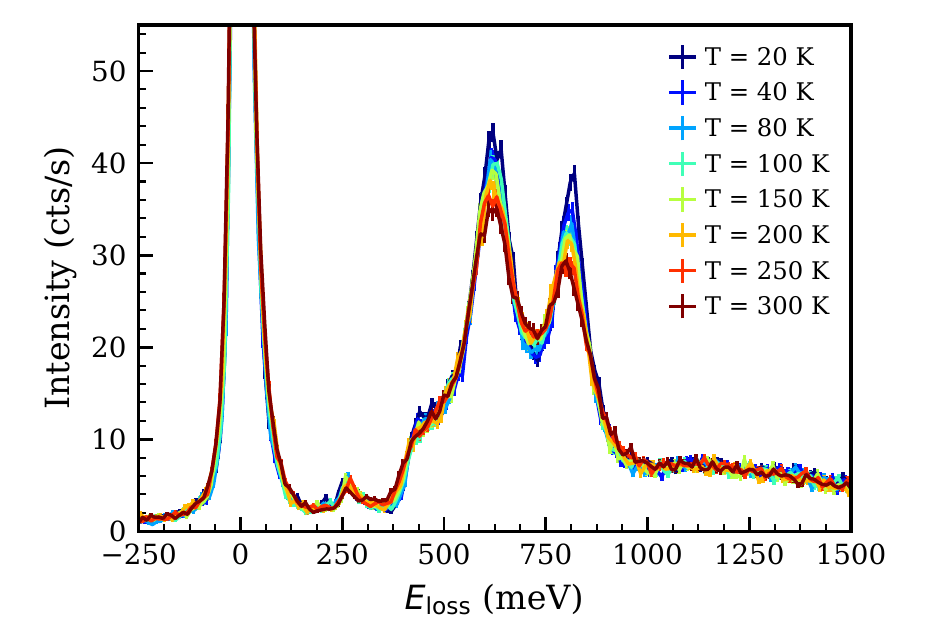}
  \caption{RIXS intensity at $E_i = 11.215$~keV as a function of temperature.}
  \label{fig:rixsT}
  \end{center}
\end{figure}

We show in Fig.~\ref{fig:rixsT} the RIXS intensity of \AgIr{} as a function of temperature. We observe a stronger temperature dependence of the intensity of peaks $D$--$E$ than that of peaks $A$--$C$, similar to that of other iridates \cite{SI_PhysRevB.95.235161}.T he broadening in energy and decrease of intensity with increasing temperature dependence may be related to Debye Waller factors \cite{SI_PhysRevB.62.13996}. This does not affect the analysis presented in the main text. 

\section{Details of EXAFS refinements}
\begin{figure}[!b]
  \begin{center}
    \includegraphics{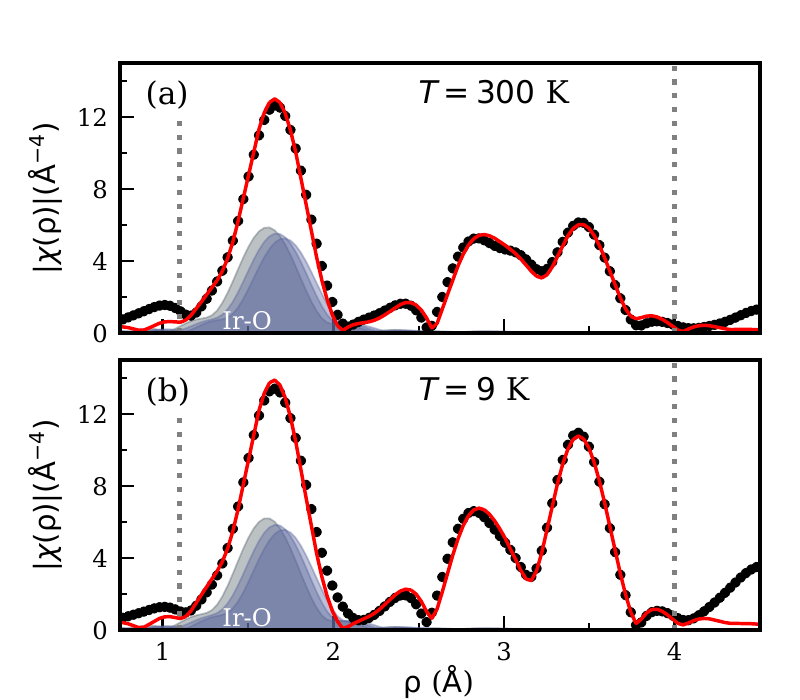}
  \caption{Magnitude of the Fourier transform of the EXAFS oscillations measured at (a) $T\! = \! 300$~K and (b) $T\! = \! 9$~K. Red line is a fit to the magnitude of the Fourier transform of $k^3\chi(k)$ in the range $\rho$=1.1--4.1 \AA{}. }
  \label{fig:exafs}
  \end{center}
\end{figure}
In Fig.~\ref{fig:exafs} (a) -- (b) we show the magnitude of the Fourier transform of the EXAFS oscillations measured at $T\! = \! 300$~K and $T\! = \! 9$~K. The red line is the magnitude of a fit to the Fourier transform of $k^3\chi(k)$. The Fourier Transform was performed in the range $k=3-15$ \AA$^{-1}$ by applying a Kaiser window of width $\Delta k = 12$ \AA$^{-1}$ with tapering $\delta k=4$ \AA$^{-1}$. The fit was then performed in real space in the range $r = 1.1-4.1$ {\AA} using least-squares minimization. The fit parameters are $E_0$, which fixes the $k=0$ point for all paths, an amplitude reduction factor $S_0^2$, which is an overall scale factor and the respective shift in path length $\delta r$ and mean-square displacement $\sigma^2$ for each each path of nominal length $r$. The fit includes single scattering paths calculated by the FEFF software package \cite{SI_REHR2009548} using the refined structure from PDF measurements as initial input. The paths include the nearest-neighbor Ir-O bonds, nearest-neighbor Ir-Ir bonds, nearest-neighbor Ir-Ag bonds and next-nearest neighbor Ir-O bonds.  We excluded the Ir-Li paths from the fit due to the large mean-square displacement of Li.

\section{Magnetic field dependence of the net moment and XMCD signal at base temperature}

\begin{figure}[!ht]
  \begin{center}
    \includegraphics{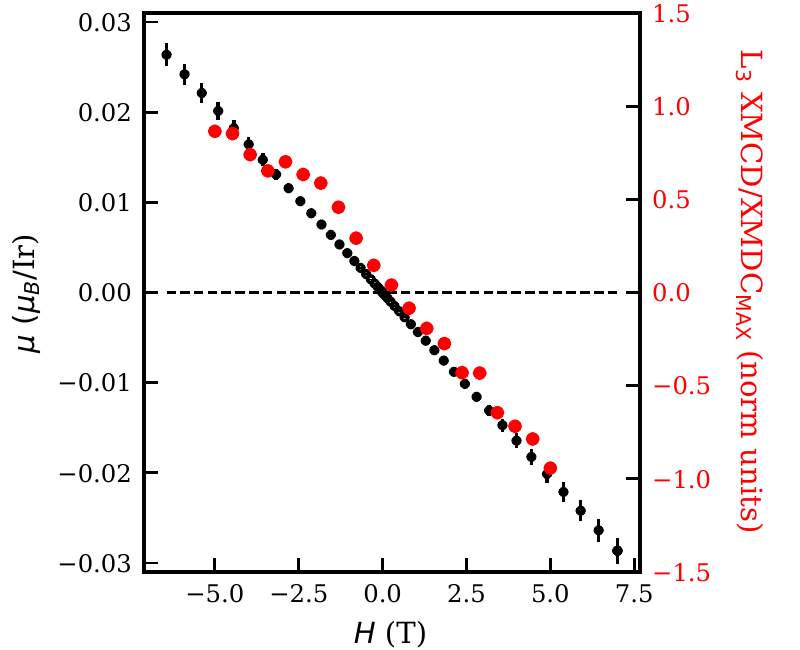}
  \caption{Net moment per Ir from bulk magnetization and L$_3$ XMCD signal at base temperature as a function of magnetic field.}
  \label{fig:M_XMCD}
  \end{center}
\end{figure}

In Fig.~\ref{fig:M_XMCD} we compare the extracted net moment from bulk magnetization measurements at $T = 2$ K and the L$_3$ XMCD signal at $T = 1.6$ K as a function of magnetic field. Both magnitudes follow a linear trend with field.

\section{Refined structural parameters from PDF and EXAFS}
We show in Table~\ref{tab:params} the refined atomic positions from the PDF analysis and in and Table~\ref{tab:paramsexafs}  local structural parameters from the PDF and EXAFS analysis.
\begin{table}[!ht]
    \centering
    \begin{tabular}{c c c c c }
    \hline
        Atom & Site & x & y & z\\
      \hline
      \hline
      Ir  & $4g$ & 0.000 & 0.334(1) & 0.000  \\
      Li & $2a$ & 0.000 & 0.000 & 0.000  \\
      O & $4i$ & 0.404(9) & 0.000 & 0.187(8)  \\
      O & $8j$ & 0.393(9) & 0.323(7) & 0.178(5)  \\
      Ag & $4h$ & 0.500 & 0.336(2) & 0.500  \\
      Ag & $2d$ & 0.500 & 0.000 & 0.500  \\
      \hline
    \end{tabular}
    \caption{Summary of structural parameters from the PDF refinement over the range $1.2<r<12$\AA{}. The room temperature cell was refined in the $C2/m$ space group to give a=5.344(5) \AA{}, b= 9.012(8) \AA{}, c= 6.468(3) \AA{} , and $\beta=105.39(8)^{\circ}$. All sites are fully occupied.}
    \label{tab:params}
\end{table}

\begin{table}[!h]
    \centering
    \begin{tabular}{c c c }
        Bond & PDF (\AA{}) & EXAFS (\AA{}) \\
      \hline
      \hline
      ${\rm Ir_1}$-${\rm O_1}$  & 2.021  & 1.954 \\
      ${\rm Ir_1}$-${\rm O_2}$  & 2.119  & 2.050 \\
      ${\rm Ir_1}$-${\rm O_3}$  & 2.074  & 2.006 \\
      ${\rm Ir_1}$-${\rm Ag_1}$ & 3.564  & 3.588 \\
      ${\rm Ir_1}$-${\rm Ag_2}$ & 3.580  & 3.605 \\
      ${\rm Ir_1}$-${\rm Ag_3}$ & 3.607  & 3.632 \\
      ${\rm Ir_1}$-${\rm Ir_2}$ & 2.994  & 3.000 \\
      ${\rm Ir_1}$-${\rm Ir_3}$ & 3.071  & 3.078 \\
    \end{tabular}
    \caption{Summary of the local structural parameters from the PDF and EXAFS analysis. The room temperature structural cell was refined in the $C2/m$ spacegroup to give a=5.343, b= 9.015, c = 6.468, $\beta=105.385$. All sites are fully occupied.}
     \label{tab:paramsexafs}
\end{table}

\section{Treatment of the trigonal fields}
For the complete basis of the spherical harmonic $(\mathcal{Y}^2_2,\mathcal{Y}^1_2,\mathcal{Y}^0_2,\mathcal{Y}^{-1}_2,\mathcal{Y}^{-2}_2$) the crystal field Hamiltonian including cubic and trigonal can be written as:

\begin{flalign}
    H^{D_{3d}}_{CF}=\begin{pmatrix}
     a_{0,0} & 0 & 0 & a_{0,3} & 0 & \\
     0 & a_{1,1} & 0 & 0 & a_{1,4} \\
     0 & 0 & a_{2,2} & 0 & 0 \\
     a_{3,0} & 0 & 0 & a_{3,3} & 0  \\
     0 & a_{4,1} & 0 & 0 & a_{4,4}  \\
    \end{pmatrix},
\end{flalign}

with:
\begin{subequations}
\begin{flalign}
&a_{0,0} = a_{4,4} = -3 C_p \mathcal{P}_1(\alpha) + (3/14) D_q
\mathcal{P}_2(\alpha),\\
&a_{0,3} = a_{3,0} = -a_{1,4} = -a_{4,1} = 15 D_q \sin{\alpha}^3 \cos{\alpha},\\
&a_{1,1} = a_{3,3} = (3/2) C_p \mathcal{P}_1(\alpha) - (6/7) D_q  \mathcal{P}_2(\alpha),\\
&a_{2,2} = 3 C_p \mathcal{P}_1(\alpha) + (9/7) D_q
\mathcal{P}_2(\alpha),
\end{flalign}
\end{subequations}
and $\mathcal{P}_1(\alpha) = 3 \cos{\alpha}^2 -1$ and $\mathcal{P}_2(\alpha) = 35 \cos{\alpha}^4 - 30 \cos{\alpha}^2 +3$. $\alpha = \arctan(\sqrt{2}\frac{Ir-O_1}{Ir-O_2})$ parameterizes the size of the trigonal distortion, with $\alpha_0 = 54.74^{\circ}$ for a perfect octahedral environment. $D_q$ encodes the cubic crystal fields and gives the splitting between $e_g$ and $t_{2g}$ levels. $C_p$ is the value of the non-cubic crystal fields due to trigonal distortion and follows $C_p = \frac{6}{35} D_q \kappa$ with $\kappa = r_0^2 \frac{\langle r^2 \rangle}{\langle r^4 \rangle}$. \textit{ab-initio} calculations give a realistic upper bound of $\kappa = 1$ \cite{SI_Khomskii2016,SI_PhysRevB.91.155125}. Notice that $\alpha = \alpha_0$ cancels the terms proportional to $C_p$.

\begin{figure}[!ht]
  \begin{center}
    \includegraphics{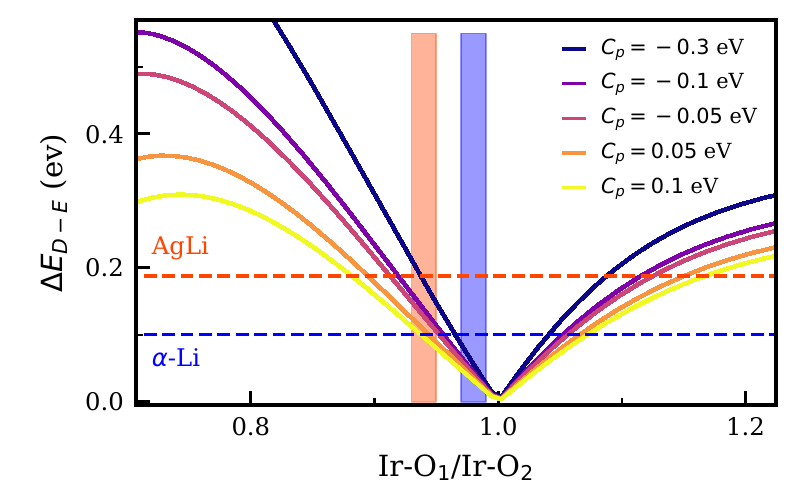}
  \caption{$\Delta E_{D-E}$ as a function of $\frac{Ir-O_1}{Ir-O_2}$ for different values of $C_p$.}
  \label{fig:cp}
  \end{center}
\end{figure}

We show in Fig.~\ref{fig:cp} dependence of $\Delta E_{D-E}$ as a function of $\frac{Ir-O_1}{Ir-O_2}$ for different values of $C_p$ keeping $D_q \!=\! 0.44$~eV and the spin-orbit coupling strength  $\lambda \!=\! 0.37$~eV constant. The observed $\Delta E_{D-E} \!=\! 0.19(4)$~eV in \AgIr{} can be reproduced for $C_p \!=\! -0.3$~eV. However, this value gives a ratio $C_p/D_q = 0.68$ that is much larger than the upper limit from \textit{ab-initio} calculations for 5$d$-electrons $C_p/D_q = 0.17$, \cite{SI_Khomskii2016,SI_PhysRevB.91.155125}. In general, the magnitude $\Delta E_{D-E}$ depends strongly on bond length disproportionation and weakly on $C_p$ and $D_q$ \cite{SI_NAGASUNDARAM1989163}.Fixing $C_p \!=\! -0.05$ eV, $C_p/D_q < 0.17$, we find that to account for the experimentally observed $D$--$E$ splitting in \Li{}, $\Delta E_{\mathrm{trig}}\!=\!0.10(3)$ eV, we need to consider an artificially large bond disproportionation, twice as large as the reported value \cite{SI_PhysRevLett.110.076402}. For \AgIr{} the observed  $D$--$E$ splitting requires a $>10\%$ Ir-O bond disproportionation, exceeding the range admitted by our local structure measurements [(Ir-O$_2$-Ir-O$_1$)/Ir-O$_1 \!=\! 5.3\%$].

In the large crystal field limit, we can consider the effects of the trigonal non-cubic crystal fields only in the $t_{2g}$ states through the Hamiltonian:
\begin{flalign}
    H^{t_{2g}}_{CF}=\begin{pmatrix}
     0 & \delta & \delta \\
     \delta & 0 &\delta\\
     \delta & \delta & 0\\
    \end{pmatrix}
\end{flalign}
in the basis ($d_{zx}$, $d_{zy}$, $d_{xy}$).

\section{Single site RIXS calculation}

\begin{figure}[!ht]
  \begin{center}
    \includegraphics{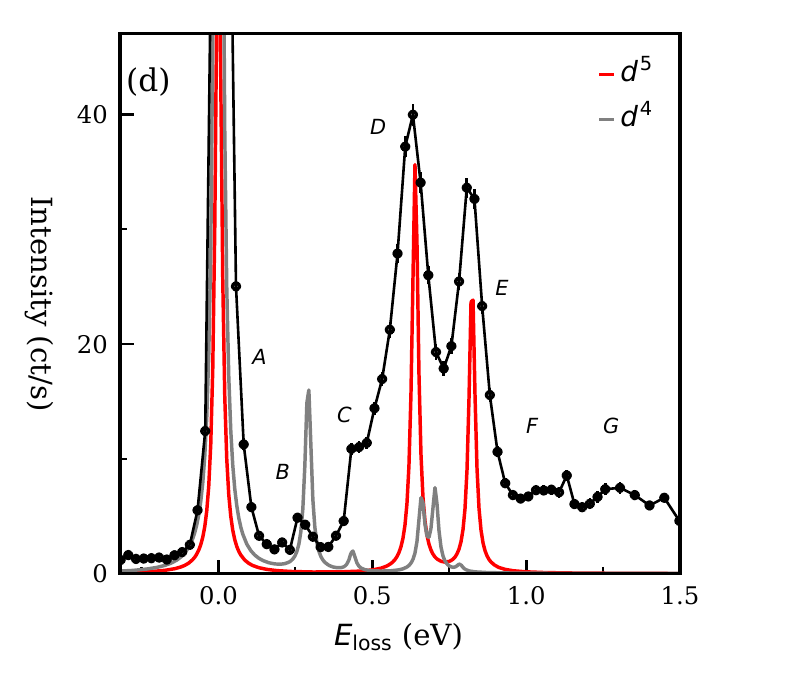}
  \caption{Calculated RIXS spectrum for Ir$^{4+} (d^5)$ and Ir$^{5+} (d^4)$ with non-cubic crystal field splitting.}
  \label{fig:d4_d5}
  \end{center}
\end{figure}

The inclusion of non-cubic crystal fields (CF) cannot reconcile the RIXS spectra shown in Fig. 3 with a single-site Ir$^{4+}$ model, which can at most produce two RIXS features for $E_{\rm loss}\! <\! 1$~eV. Indeed, density functional theory calculations have shown that the effects of trigonal distortions on RIXS spectra in the honeycomb iridates are overestimated when direct Ir-Ir hopping terms are neglected \cite{SI_PhysRevLett.109.197201,SI_PhysRevB.88.035107,SI_PhysRevB.78.012501}. To demonstrate this, we show in Fig.~\ref{fig:d4_d5} calculations for a single site Ir $5d^5$ and $5d^4$ ion in a trigonally distorted environment given by $H^{D_{3d}}_{CF}$ with $10Dq\!=\!4.4$ eV and $C_p \!=$-0.05 eV, including spin orbit coupling $\lambda$. We find the best agreement with the experimental RIXS data for $\lambda = 0.37$~eV. Electron-electron correlations are parameterized using Slater integrals calculated using Cowan's code \cite{SI_Cowan,SI_WANG2019151}, reported in Table~\ref{tab:slater}.  The on-site Coulomb repulsion between $5d$ electrons is given by the integrals $F^0_{dd}$,$F^2_{dd}$, and $F^4_{dd}$. The Coulomb interaction between the $2p$-core hole and $5d$ electrons is given by $F^0_{pd}$, $F^2_{pd}$, $G^1_{pd}$ and $G^3_{pd}$.

\begin{table}[!h]
    \centering
    \begin{tabular}{c c c }
      
        & $5d^5$ & $5d^4$ \\
      \hline
      \hline
      $F^0_{dd}$ & 0.47 & 0.49 \\
      $F^2_{dd}$ & 8.83 & 9.20 \\
      $F^4_{dd}$ &5.91 & 6.18 \\
      $F^0_{pd}$ &0.08&0.09 \\
      $F^2_{pd}$ &1.03& 1.07\\
      $G^1_{pd}$ &0.89 & 0.96\\
      $G^3_{pd}$ & 0.53& 0.57\\
    \end{tabular}
    \caption{Summary of Slater parameters used in the single site RIXS calculations including all $5d$ orbitals.}
     \label{tab:slater}
     
\end{table}

These calculations for both $d^4$ and $d^5$ do not reproduce features $C$,$F$ and $G$, reaffirming the need for considering Ir-Ir hopping terms. Our RIXS calculations in Fig.~\ref{fig:d4_d5} for a Ir$^{5+}$ ion show an inelastic feature that coincides in energy with peak $B$, $B\!=\!0.270(37)$ eV. In Section S6 we ruled out Ir$^{5+}$ impurities concentrations $>1\%$ to be the origin of this feature in \AgIr{}. Additionally, this model does not properly capture the relative intensities of features $B$ and $C$. 

\begin{figure*}[!ht]
  \begin{center}
    \includegraphics[width=1\textwidth,clip]{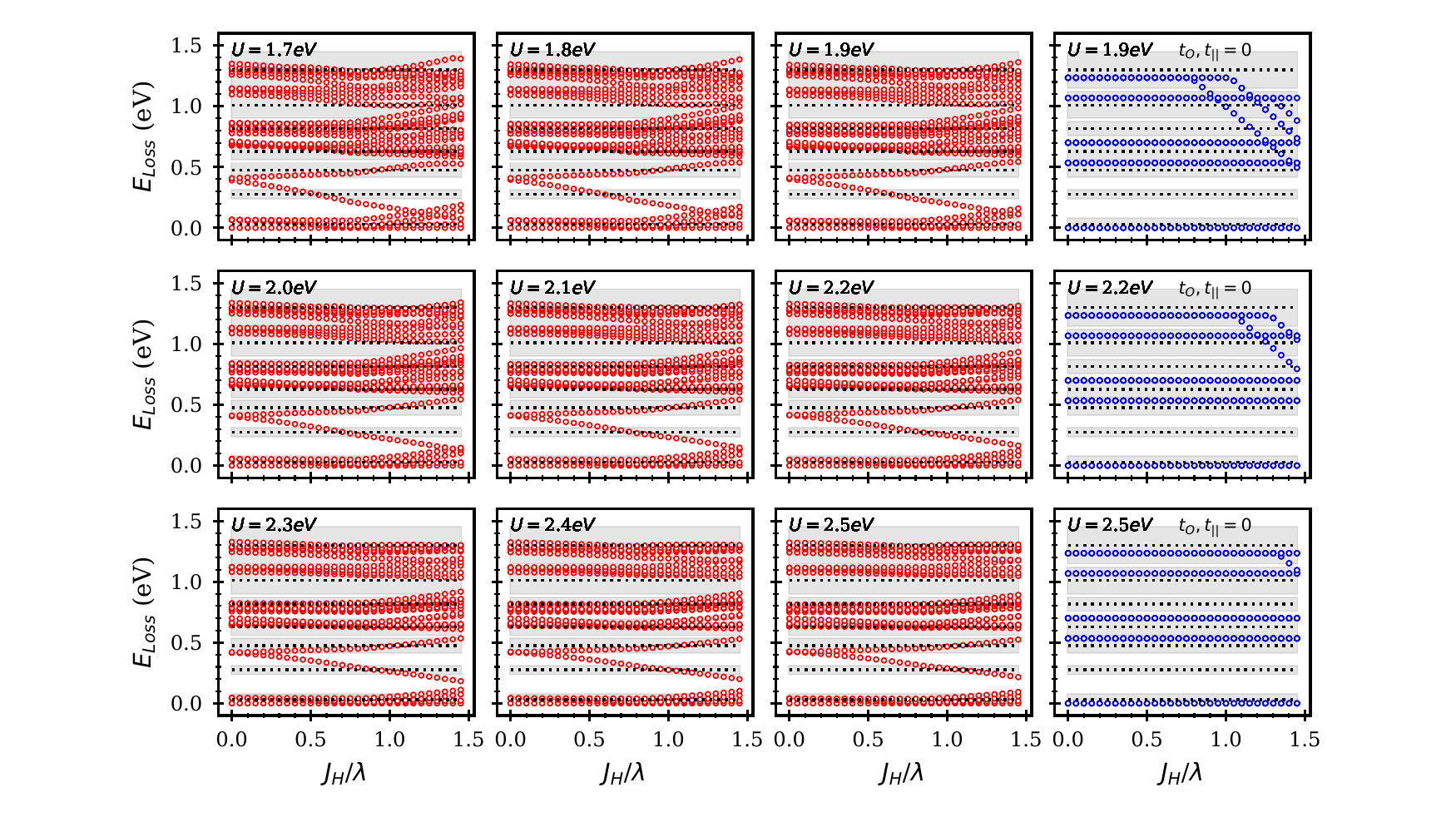}
  \caption{Result of the exact diagonalization calculations as a function of $J_H$ and $U$, with $\lambda = 0.395$ eV. Red markers indicate calculated spectrum with $\delta = -0.075$ eV, $t_{\parallel} = -0.085 $ eV, $t_{\rm O} = 0.525 $ eV , $t_{\sigma}\!=\!-t_{\parallel}$, and $t_{\perp} \!=\! -0.05 t_{\parallel}$. For the calculations indicated by blue markers all the hopping terms are set to zero.}
  \label{fig:UJH}
  \end{center}
\end{figure*}

\section{Exact diagonalization including hopping integrals}
We consider a model for the low energy electronic structure of two Ir sites in the large cubic crystal field limit given by the Hamiltonian $H = H_U + V_{12}$. The Kanamori-type Hamiltonian $H_U$ is given by
\begin{flalign}
\begin{split}
    H_U=\\
    &U\sum_{\psi}n_{\psi\uparrow} n_{\psi\downarrow}+ U'\sum_{\psi \neg \psi'}n_{\psi\uparrow} n_{\psi'\downarrow}\\
    &+U'\sum_{\psi < \psi', \sigma}n_{\psi\sigma} n_{\psi'\sigma}-J_H \sum_{\psi \neg \psi'} c^{\dagger}_{\psi\uparrow}c_{\psi\downarrow}c^{\dagger}_{\psi'\downarrow}c_{\psi'\uparrow}\\
    &+J_H \sum_{\psi \neg \psi'} c^{\dagger}_{\psi\uparrow}c^{\dagger}_{\psi\downarrow}c_{\psi'\downarrow}c_{\psi'\uparrow},
\end{split}
\end{flalign}
with on-site Coulomb interaction, $U \!=\! 2$ eV, and Hunds's coupling, $J_H \!=\! 0.3$ eV, \cite{SI_PhysRevLett.122.106401}. The hopping between Ir sites, $V_{12}$ is given by

\begin{equation}
    V_{12}=\begin{pmatrix}
     \lambda & \delta & \delta & t_{\parallel} & t_{\rm O} & t_{\perp}\\
     \delta & \lambda & \delta & t_{\rm O} & t_{\parallel} & t_{\perp}\\
     \delta &  \delta & \lambda & t_{\perp} & t_{\perp} & t_{\sigma}\\
     t_{\parallel} & t_{\rm O} & t_{\perp} &  \lambda & \delta & \delta \\
     t_{\rm O} & t_{\parallel} & t_{\perp} &  \delta & \lambda & \delta \\
      t_{\perp} & t_{\perp} & t_{\sigma} &  \delta &  \delta & \lambda\\
    \end{pmatrix},
\end{equation}

for a complete $d$-orbital basis ($d_{z_1x_1}$, $d_{z_1y_1}$, $d_{x_1y_1}$, $d_{z_2x_2}$, $d_{z_2y_2}$, $d_{x_2y_2}$) following following the labels and nearest-neighbour paths previously considered to to model quasi-molecular orbitals \cite{SI_PhysRevLett.109.197201,SI_PhysRevB.88.035107,SI_PhysRevMaterials.4.075002,SI_Streltsov_2017}.  $\delta$ is the non-cubic crystal field strength  and we fix the spin orbit coupling constant $\lambda \!=\! 0.395 $ eV.

\section{Role of $J_H$ and $U$}

To efficiently explore the set of energy scales that affect the low energy electronic structure of \AgIr{}, we fixed the on-site electron-electron correlation, $U \!=\! 2$ eV, and Hund's coupling $J_H \!=\! 0.3$ eV. These values are well within the range of values used in the literature for other iridates ($ 1.7 < U < 2.5$ eV, and $0.2 < J_H < 0.5$ eV) \cite{SI_PhysRevB.97.235119,SI_PhysRevB.98.245123,SI_PhysRevB.95.235114,SI_PhysRevB.93.214431}. In Fig.~\ref{fig:UJH}, we show the result of the exact diagonalization calculations as a function of $J_H$ and $U$, with $\lambda \!=\! 0.395 $ eV and $\delta = -0.075$ eV. For $t_{\parallel} = -0.085 $ eV, $t_{\rm O} = 0.525 $ eV , $t_{\sigma}\!=\!-t_{\parallel}$, and $t_{\perp} \!=\! -0.05 t_{\parallel}$, red markers, we observe minimal variation of the calculated spectrum across a wide range of values of $U$. On the other hand, the splitting between calculated energy levels, in particular below $E_{loss} < 0.5$ eV, depends linearly on $J_H/\lambda$. The dashed horizontal lines and shaded boxes indicate the centroids and FWHM of peaks $A$ -- $G$. This highlights the dependence of the low energy electronic structure in $4+$ iridates on multiple energy scales. When the hopping integrals are not considered, $V_{1,2} = 0$, blue markers, the electronic spectrum cannot account for the observed RIXS dispersion for any physical value of $J_H$ and $U$.

\section{Quality Factor}

\begin{figure*}[!ht]
  \begin{center}
    \includegraphics[width=0.9\textwidth,clip]{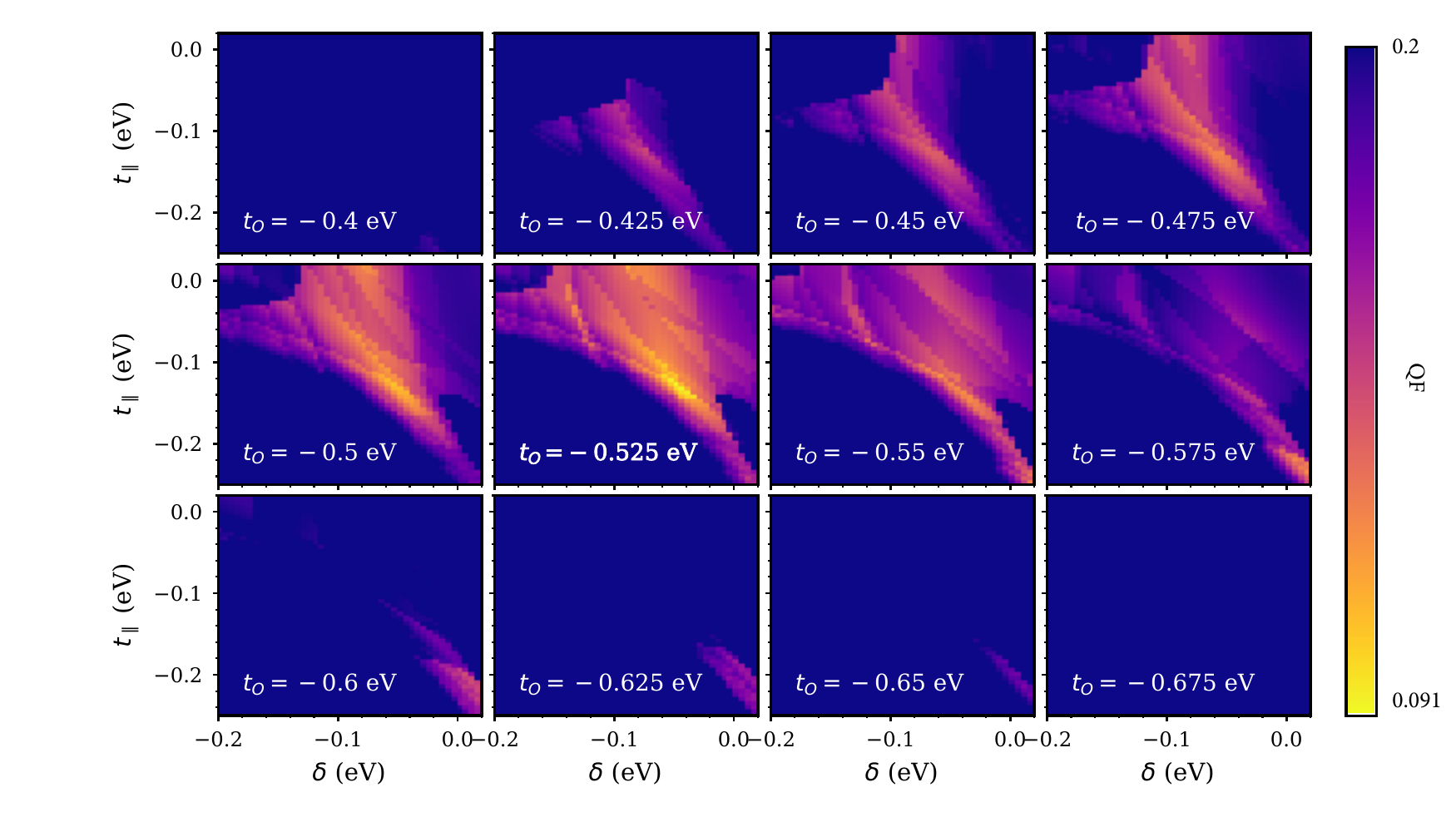}
  \caption{The false color map encodes the value of QF as a function of $t_{\parallel}$, $t_{O}$ and $\delta$.}
  \label{fig:QF}
  \end{center}
\end{figure*}

To efficiently explore the phase space defined by the hopping integrals and the non-cubic crystal fields, we define a \textit{quality factor} QF to parameterize the agreement between measured and calculated exciton energies
\begin{equation}
    QF =\sum_{p \in \{A,B, C,D,E\}} \frac{1}{N_0} \sum_{i=0}^{N_0}\sqrt{(E_0 - E_{p})^2}/{\sigma}_{p}.
\end{equation}
$E_p$ is the measured energy of peak $p\!\in\!\{A,B,C,D,E\}$ with FWHM of $\sigma_p$ [see Fig. 1, main text], $N_{0}$ is the number of eigenvalues that satisfy $[E_0-\delta E,E_0+\delta E] \in [E_{p}-\sigma/2,E_{p}+\sigma/2]$, with $\delta E = 0.033$~eV the experimental resolution. Any eigenvalues that lie within range of more than one experimental peak are assigned to the lowest energy peak of the set. We exclude all calculated states at energy transfers less than $\Delta E/2$ from the QF calculation to avoid complications from the elastic line. Experimental features $F$, $G$ and $H$, with \eloss{}$> 1$~eV and proximal to the $e_g$ features, are also excluded from the calculation as their broad energy width artificially enhances their weights and biases the calculation of QF.

We first assume $t_{\sigma}\!=\!-t_{\parallel}$, and $t_{\perp} \!=\! -0.05 t_{\parallel}$. We then diagonalize the Hamiltonian for different values of $t_{\parallel}$, $t_{O}$ and $\delta$ in order to find the parameter set that minimizes QF. We perform a brute force minimization in order to explore any correlations between parameters and constrain a global minimum for QF. In Fig.~\ref{fig:QF} we show the dependence of QF on $t_{\parallel}$, $t_{O}$ and $\delta$ over a broad range of physically realistic values. Our QF map reveal that $t_{\parallel}$ and $\delta$ are highly correlated: the minimum of QF cannot be reached by varying both parameters independently and for larger $t_{\parallel}$ QF is minimized with smaller $\delta$.

In Fig.~\ref{fig:rixscomp}, we show a comparison of our data with calculated powder averaged RIXS spectra for a range of parameters that fall withing 10\% of the minimum QF. 

\section{Powder Averaged RIXS cross-section calculation}
We use EDRIXS software package to diagonalize the full Hamiltonian $H = H_U + V_{12}$ and simulate the full momentum and energy dependent RIXS spectra within a dipole approximation \cite{SI_WANG2019151}. We consider a cluster of four Ir atoms and average the RIXS signal over the three inequivalent Ir-Ir bonds within a Li$_2$IrO$_3$ layer. In our calculation, we set $2\theta = 90^{\circ}$, fix the incident beam polarization $\bm{\epsilon}_i$ to lie in the horizontal scattering plane and average over all outgoing polarization directions $\bm{\epsilon}_f$ to match the experimental geometry. 

The RIXS cross-section is given by
\begin{flalign}
\begin{split}
    &I_{\rm RIXS} (\omega_{\rm in},\omega_{\rm loss},\bm{k}_i,\bm{k}_f,\bm{\epsilon}_i,\bm{\epsilon}_f) = \sum\limits_{i} \frac{1}{Z} \exp{\left[-\frac{E_i}{k_B T}\right]} \times\\
    &\sum\limits_n \bigg\rvert \frac{ \langle f \rvert \mathcal{\hat{D}}^{\dagger}_f\rvert n\rangle \langle n \rvert \mathcal{\hat{D}}_i \rvert i\rangle}{\omega_{\rm in}-\hat{H}_n-E_i+i\Gamma_c}\bigg\rvert^2 \frac{\Gamma/\pi}{(\omega_{\rm loss}-E_{f}+E_i)^2+\Gamma^2},
    \label{eq:crosssection}
\end{split}
\end{flalign}
where  $\bm{k}_i$, $\bm{k}_f$, $\omega_{\rm in}$, and  $\omega_{\rm loss}$ are the incident and outgoing x-ray wavevectors, incident x-ray energy, and energy transfer respectively. $\rvert i \rangle$ and $\rvert f \rangle$ correspond to the eigenvalue of the initial and final states, with $E_i$ and $E_f$ the corresponding eigenvalues. $Z$ is the partition function and $k_B$ the Boltzmann constant. $\rvert n \rangle$ corresponds to eigenstates of the intermediate state Hamiltonian which includes a $2p$ core hole and $n+1$ electrons in the valence shell after absorbing a photon with energy $\omega_{\rm in}$. $\Gamma_c$ and $\Gamma$ are the lifetime broadening of the intermediate and final state.

$\hat{D}_i$ and $\hat{D}^{\dagger}_f$ are the transition operators for the x-ray absorption and emission processes. We only include dipolar transitions in our calculation. In this case:
\begin{subequations}
\begin{align}
     &\mathcal{\hat{D}}_{i} = \frac{1}{im\omega_{\rm in}}\sum_{i=1}^{N}e^{i\bm{k}_i\cdot\bm{R}_{\alpha}}\bm{\epsilon}_i^{}\,{\cdot}\,\bm{p}_{\alpha}^{}, \\
     &\mathcal{\hat{D}^{\dagger}}_f = \frac{1}{-im\omega_{f}}\sum_{i=1}^{N}e^{-i\bm{k}_f\cdot\bm{R_{\alpha}}}\bm{\epsilon}_f^{}\,{\cdot}\,\bm{p}^{\dagger}_{\alpha},
\end{align}
\end{subequations}
where $\bm{R}_{\alpha}$ is the position of the resonant ion to which electron ${\alpha}$ is bound and $\bm{p}_{\alpha}$ is the momentum operator. The term $\mathcal{\hat{D}}$ encodes the momentum dependence and depends on the direction between x-ray polarization and the position (momentum) of the bounded electron. For a powder sample, the measured intensity will be a spherical average of this term at fixed momentum transfer.

\begin{figure*}[!ht]
  \begin{center}
    \includegraphics[width=1\textwidth,clip]{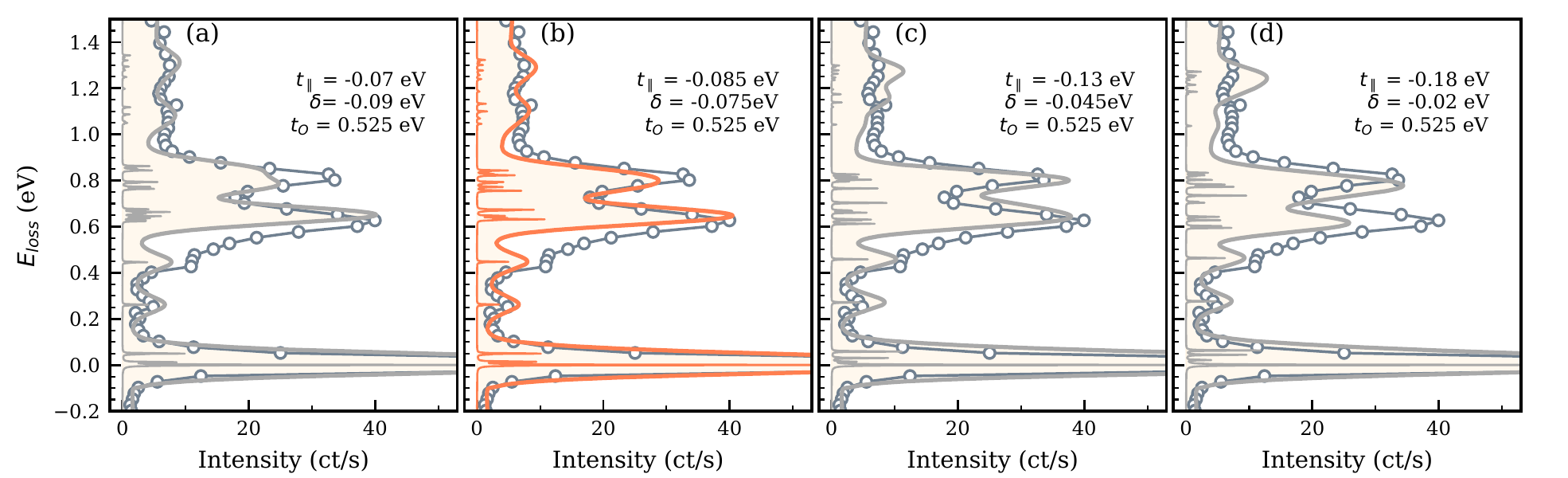}
  \caption{Comparison of the calculated RIXS intensity for different values of $t_{\parallel}$, $t_{O}$ and $\delta$ that are within $10\%$ of the QF minimum. Panel (b), highlighted in orange, shows the parameter set from QF minimization that best matches the RIXS spectrum and is presented in Fig. 4 of the main text.}
  \label{fig:rixscomp}
  \end{center}
\end{figure*}

In order to compare the calculated RIXS intensity with our powder measurement, we perform powder averaging by Monte-Carlo (N = 1000) integration of equation \ref{eq:crosssection} over a spherical surface of radius $|Q|=k_i\sin{\left(2\theta/2\right)}$. Although the calculated RIXS excitations from our two site model do not show any dispersion, the powder averaging affects the relative intensities of RIXS features through the transition operators. For our calculations we fixed $\Gamma = 0.005$~eV and we account for experimental resolution through a convolution of the calculated RIXS intensity with a Gaussian profile of FWHM ~ 33~eV. The results of this calculation using a range of  optimal parameters set determined by our QF minimization is shown in Fig.~\ref{fig:rixscomp} and Fig.~4 (d) of the main text.

\end{document}